\documentclass[preprint2]{aastex6}

\setcitestyle{notesep={ }}

\newcommand\corot{{\it CoRoT}}
\newcommand\kepler{{\it Kepler}}
\newcommand\ktwo{{\it K2}}

\newcommand\kms{\hbox{km\,s$^{-1}$}}  
\newcommand\ms{\hbox{m\,s$^{-1}$}}  

\newcommand\vsini{$v$\,sin\,$i_\star$}   
\newcommand\vmic{$v_{\rm mic}$}
\newcommand\vmac{$v_{\rm mac}$}
\newcommand\teff{$T_{\rm eff}$}
\newcommand\logg{log\,{\it g$_\star$}}

\newcommand{\smass}[1][$M_{\odot}$]{ $ 0.877 \pm 0.024 $~#1} 
\newcommand{\sradius}[1][$R_{\odot}$]{ $0.835 \pm 0.026$~#1}

\newcommand{\Tzerob}[1][days]{$7394.37442 _{ - 0.00055 } ^ { + 0.00060 } $#1} 
\newcommand{\Pb}[1][days]{$0.959632 \pm 0.000015 $ #1}  
\newcommand{\bb}[1][]{$0.11 _{ - 0.08 } ^ { + 0.11 }$ #1}       
\newcommand{\arb}[1][]{$4.516 _{ - 0.085 } ^ { + 0.076 }$ #1}       
\newcommand{\rrb}[1][]{$0.01728 \pm 0.00025 $ #1}       
\newcommand{\kb}[1][$m \, s^{-1}$]{$4.02 \pm 0.31$~#1} 
\newcommand{\ib}[1][deg]{$ 88.6 _{ - 1.4 } ^ {+ 1.0 }$ #1} 
\newcommand{\ab}[1][AU]{$ 0.01752 \pm 0.00063 $ #1}   
\newcommand{\mpb}[1][$M_{\oplus}$]{$5.69 \pm 0.44 $~#1} 
\newcommand{\rpb}[1][$R_{\oplus}$]{$1.574 \pm 0.054 $~#1}   
\newcommand{\denpb}[1][$\mathrm{g\,cm^{-3}}$]{$8.00 _{ - 0.98 } ^ { + 1.10  }$~#1}
\newcommand{\Tequib}[1][K]{$ 1759 \pm 20 $  #1}    
\newcommand{\ttotb}[1][hours]{$ 1.65 \pm 0.03 $  #1}

\newcommand{\Tzeroc}[1][days]{$7394.97831 \pm 0.00085 $#1} 
\newcommand{\Pc}[1][days]{$29.84622 _{ - 0.00091 } ^ { + 0.00098  }$ #1} 
\newcommand{\esinc}[1][]{$0.00 _{ - 0.24 } ^ { + 0.17  }$ #1 } 
\newcommand{\ecosc}[1][]{$0.06 _{ - 0.17 } ^ { + 0.16  }$ #1 } 
\newcommand{\ec}[1][]{$0.05 _{ - 0.04 } ^ { + 0.07 }$ #1}      
\newcommand{\wc}[1][deg]{$178 _{ - 136 } ^ { + 134 }$ #1} 
\newcommand{\bc}[1][]{$0.30 _{ - 0.18 } ^ { + 0.11 }$ #1}       
\newcommand{\arc}[1][]{$46.5 \pm 1.5 $ #1}       
\newcommand{\rrc}[1][]{$0.03006 _{ - 0.00055 } ^ { + 0.00065  }$ #1}       
\newcommand{\kc}[1][$m \, s^{-1}$]{$1.88_{-0.42}^{+0.40}$~#1} 
\newcommand{\ic}[1][deg]{$ 89.6 \pm 0.2 $ #1} 
\newcommand{\ac}[1][AU]{$ 0.1806 \pm 0.0080 $ #1}   
\newcommand{\mpc}[1][$M_{\oplus}$]{$8.33 _{ - 1.85  } ^ {+ 1.79 }$~#1} 
\newcommand{\rpc}[1][$R_{\oplus}$]{$2.740 _{ - 0.100  } ^ {+ 0.106 }$~#1}
\newcommand{\denpc}[1][$\mathrm{g\,cm^{-3}}$]{$2.21 _{ - 0.53 } ^ { + 0.56  }$~#1}
\newcommand{\Tequic}[1][K]{$ 548 \pm 10 $  #1}    
\newcommand{\ttotc}[1][hours]{$ 4.81 _{ - 0.09  } ^ {+ 0.17 }$  #1}

\newcommand{\Pd}[1][days]{$10.77 _{ - 0.13 } ^ { + 0.15  }$ #1}      
\newcommand{\kd}[1][$m \, s^{-1}$]{$1.34 _{ - 0.28  } ^ {+ 0.27 }$ #1} 
\newcommand{\mpd}[1][$M_{\oplus}$]{$4.24 _{ - 0.89  } ^ {+ 0.87 }$ #1} 
 
\newcommand{\Pe}[1][days]{$5.967 _{ - 0.035 } ^ { + 0.038  }$ #1}       
\newcommand{\ke}[1][$m \, s^{-1}$]{$1.26 \pm 0.25 $ #1} 
\newcommand{\mpe}[1][$M_{\oplus}$]{$3.28 \pm 0.65 $ #1} 
\newcommand{\qone}[1][]{ $0.34_{ - 0.15}^{ + 0.26 } $ #1}   
\newcommand{\qtwo}[1][]{ $0.47_{ - 0.22}^{ + 0.29 } $ #1}   
\newcommand{\uone}[1][]{ $0.54_{ - 0.17}^{ + 0.15 } $ #1}   
\newcommand{\utwo}[1][]{ $0.04_{ - 0.27}^{ + 0.35 } $ #1}   
\newcommand{\velHARPSN}[1][$\mathrm{km\,s^{-1}}$]{ $ 19.51471 \pm 0.00036 $ #1}
\newcommand{\velHARPS}[1][$\mathrm{km\,s^{-1}}$] { $ 19.52311 \pm 0.00029 $ #1}
\newcommand{\rvjitterN}[1][$\mathrm{m\,s^{-1}}$]{ $ 0.95_{ - 0.20}^{ + 0.24 } $ #1}    
\newcommand{\rvjitterH}[1][$\mathrm{m\,s^{-1}}$]{ $ 1.44_{ - 0.21}^{ + 0.24 } $ #1}

\begin{document}




\shortauthors{Gandolfi et al.}
\shorttitle{Mass determinations of HD\,3167\,b and HD\,3167\,c}

\title{The transiting multi-planet system HD\,3167: \\ a 5.7~$M_\oplus$ Super-Earth and a 8.3~$M_\oplus$ mini-Neptune}

\author{Davide~Gandolfi\altaffilmark{1}, 
Oscar~Barrag\'an\altaffilmark{1}, 
Artie~P.~Hatzes\altaffilmark{2},
Malcolm~Fridlund\altaffilmark{3,4}, 
Luca~Fossati\altaffilmark{5},
Paolo~Donati\altaffilmark{6}, 
Marshall~C.~Johnson\altaffilmark{7},
Grzegorz~Nowak\altaffilmark{8,9}, 
Jorge~Prieto-Arranz\altaffilmark{8,9}, 
Simon~Albrecht\altaffilmark{10},
Fei~Dai\altaffilmark{11,12},
Hans~Deeg\altaffilmark{8,9}, 
Michael~Endl\altaffilmark{13},
Sascha~Grziwa\altaffilmark{14},
Maria~Hjorth\altaffilmark{10},
Judith~Korth\altaffilmark{14}, 
David~Nespral\altaffilmark{8,9}, 
Joonas~Saario\altaffilmark{15},
Alexis~M.\,S.~Smith\altaffilmark{16},
Giuliano~Antoniciello\altaffilmark{1},
Javier~Alarcon\altaffilmark{17},
Megan~Bedell\altaffilmark{18},
Pere~Blay\altaffilmark{8,15},
Stefan~S.~Brems\altaffilmark{19},
Juan~Cabrera\altaffilmark{16}, 
Szilard~Csizmadia\altaffilmark{16},
Felice~Cusano\altaffilmark{20},
William~D.~Cochran\altaffilmark{13}, 
Philipp~Eigm\"uller\altaffilmark{16}, 
Anders~Erikson\altaffilmark{16}, 
Jonay~I.~Gonz\'alez Hern\'andez\altaffilmark{8,9}, 
Eike~W.~Guenther\altaffilmark{2}, 
Teruyuki~Hirano\altaffilmark{21}, 
Alejandro~S.~Mascare\~no\altaffilmark{8,22},
Norio~Narita\altaffilmark{23,24,25}, 
Enric~Palle\altaffilmark{8,9}, 
Hannu~Parviainen\altaffilmark{8,9},
Martin~P\"atzold\altaffilmark{14}, 
Carina~M.~Persson\altaffilmark{4},
Heike~Rauer\altaffilmark{16,26},
Ivo~Saviane\altaffilmark{17},
Linda~Schmidtobreick\altaffilmark{17},
Vincent~Van~Eylen\altaffilmark{3},
Joshua~N.~Winn\altaffilmark{11},
Olga~V.~Zakhozhay\altaffilmark{27}
}

\altaffiltext{1}{Dipartimento di Fisica, Universit\'a di Torino, via P. Giuria 1, 10125 Torino, Italy; email: davide.gandolfi@unito.it}
\altaffiltext{2}{Th\"uringer Landessternwarte Tautenburg, Sternwarte 5, D-07778 Tautenberg, Germany}
\altaffiltext{3}{Leiden Observatory, University of Leiden, PO Box 9513, 2300 RA, Leiden, The Netherlands}
\altaffiltext{4}{Department of Earth and Space Sciences, Chalmers University of Technology, Onsala Space Observatory, 439 92 Onsala, Sweden}
\altaffiltext{5}{Space Research Institute, Austrian Academy of Sciences, Schmiedlstrasse 6, 8042, Graz, Austria}
\altaffiltext{6}{INAF - Osservatorio Astrofisico di Arcetri, Largo E. Fermi 5, 50125, Florence, Italy}
\altaffiltext{7}{Department of Astronomy, The Ohio State University, 140 West 18th Ave., Columbus, OH 43210, USA}
\altaffiltext{8}{Instituto de Astrof\'\i sica de Canarias, C/\,V\'\i a L\'actea s/n, 38205 La Laguna, Spain}
\altaffiltext{9}{Departamento de Astrof\'isica, Universidad de La Laguna, 38206 La Laguna, Spain}
\altaffiltext{10}{Stellar Astrophysics Centre, Department of Physics and Astronomy, Aarhus University, Ny Munkegade 120, DK-8000 Aarhus C, Denmark}
\altaffiltext{11}{Department of Astrophysical Sciences, Princeton University, 4 Ivy Lane, Princeton, NJ 08544, USA}
\altaffiltext{12}{Department of Physics and Kavli Institute for Astrophysics and Space Research, Massachusetts Institute of Technology, Cambridge, MA 02139, USA}
\altaffiltext{13}{Department of Astronomy and McDonald Observatory, University of Texas at Austin, 2515 Speedway,~Stop~C1400,~Austin,~TX~78712,~USA}
\altaffiltext{14}{Rheinisches Institut f\"ur Umweltforschung an der Universit\"at zu K\"oln, Aachener Strasse 209, 50931 K\"oln, Germany}
\altaffiltext{15}{Nordic Optical Telescope, Apartado 474, 38700, Santa Cruz de La Palma, Spain}
\altaffiltext{16}{Institute of Planetary Research, German Aerospace Center, Rutherfordstrasse 2, 12489 Berlin, Germany}
\altaffiltext{17}{European Southern Observatory, Alonso de Cordova 3107, Santiago, Chile}
\altaffiltext{18}{Department of Astronomy and Astrophysics, University of Chicago, 5640 S. Ellis Ave, Chicago, IL 60637, USA}
\altaffiltext{19}{Landessternwarte K\"onigstuhl, Zentrum f\"ur Astronomie der Universit\"at Heidelberg, K\"onigstuhl 12, D-69117 Heidelberg}
\altaffiltext{20}{INAF - Osservatorio Astronomico di Bologna, Via Ranzani, 1, 20127, Bologna, Italy}
\altaffiltext{21}{Department of Earth and Planetary Sciences, Tokyo Institute of Technology, 2-12-1 Ookayama, Meguro-ku, Tokio 152-8551, Japan}
\altaffiltext{22}{Observatoire Astronomique de l'Universit\'e de Gen\`eve, 1290 Versoix, Switzerland}
\altaffiltext{23}{Department of Astronomy, The University of Tokyo, 7-3-1 Hongo, Bunkyo-ku, Tokyo 113-0033, Japan}
\altaffiltext{24}{Astrobiology Center, NINS, 2-21-1 Osawa, Mitaka, Tokyo 181-8588, Japan}
\altaffiltext{25}{National Astronomical Observatory of Japan, NINS, 2-21-1 Osawa, Mitaka, Tokyo 181-8588, Japan}
\altaffiltext{26}{Center for Astronomy and Astrophysics, TU Berlin, Hardenbergstr. 36, 10623 Berlin, Germany}
\altaffiltext{27}{Main Astronomical Observatory, National Academy of Sciences of the Ukraine, 27 Akademika Zabolotnoho St. 03143, Kyiv, Ukraine} 

\begin{abstract}
HD\,3167 is a bright (V=8.9 mag) K0\,V star observed by the NASA's {\it K2} space mission during its Campaign~8. It has been recently found to host two small transiting planets, namely, HD\,3167\,b, an ultra short period (0.96~d) super-Earth, and HD\,3167\,c, a mini-Neptune on a relatively long-period orbit (29.85~d). Here we present an intensive radial velocity follow-up of HD\,3167 performed with the FIES@NOT, HARPS@ESO-3.6m, and HARPS-N@TNG spectrographs. We revise the system parameters and determine radii, masses, and densities of the two transiting planets by combining the \ktwo\ photometry with our spectroscopic data. With a mass of \mpb, radius of \rpb, and mean density of \denpb, HD\,3167\,b joins the small group of ultra-short period planets known to have a rocky terrestrial composition. HD\,3167\,c has a mass of \mpc\ and a radius of \rpc, yielding a mean density of \denpc, indicative of a planet with a composition comprising a solid core surrounded by a thick atmospheric envelope. The rather large pressure scale height ($\sim$350\,km) and the brightness of the host star make HD\,3167\,c an ideal target for atmospheric characterization via transmission spectroscopy across a broad range of wavelengths. We found evidence of additional signals in the radial velocity measurements but the currently available data set does not allow us to draw any firm conclusion on the origin of the observed variation.
\end{abstract}





\section{Introduction} 
\label{Sect:Introduction}

Back in 1995 the discovery of \object{51\,Peg\,b} demonstrated that gas-giant planets ($R_\mathrm{p}$\,$\approx$\,1\,$R_\mathrm{Jup}$) could have orbital periods of a few days and thus exist quite close to their host star \citep{Mayor1995}. Space-based transit search missions such as \corot\ \citep{Baglin2006}, \kepler\ \citep{Borucki2010}, and \ktwo\ \citep{Howell2014} have established that these close-in planets can have radii down to Neptune-like \citep{Barragan2016,David2016} and even Earth-like values \citep{Queloz2009,Howard2009,Pepe2013}. Close-in exoplanets have challenged planet formation theories and play an important role in the architecture of exoplanetary systems \citep[e.g.,][]{Winn2015,Hatzes2016}.

\begin{figure*}
\includegraphics[width=\textwidth]{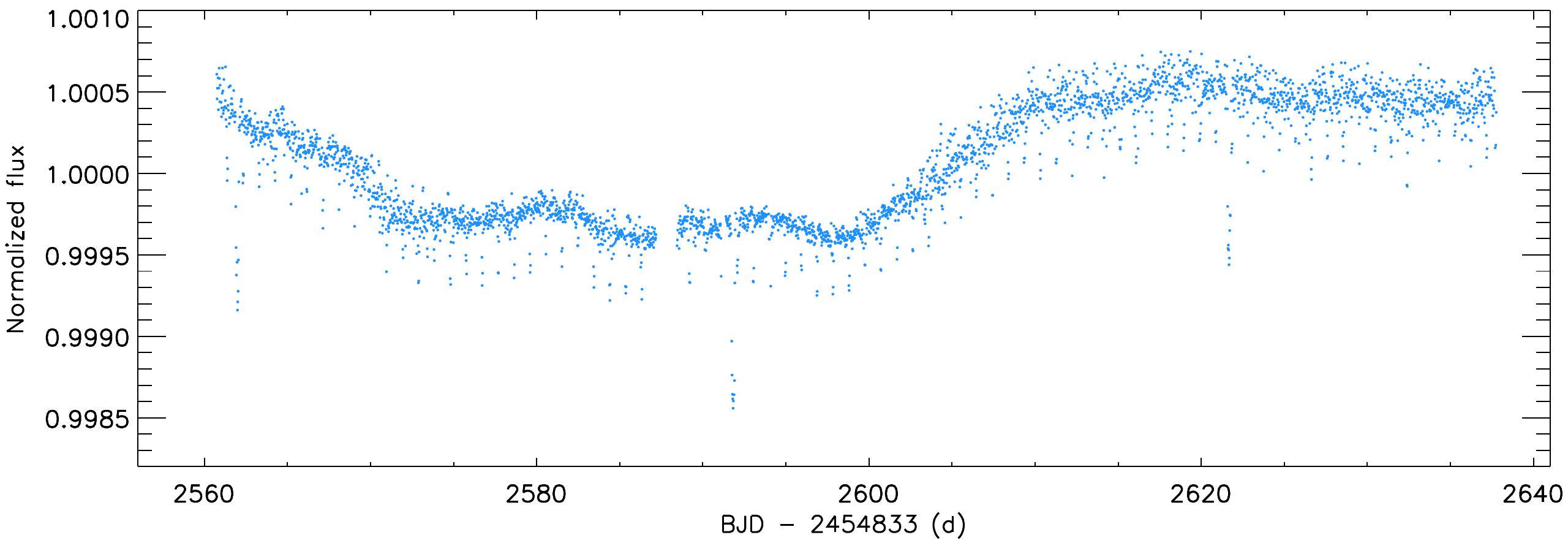}
\caption{\ktwo\ light curve of HD\,3167 from \citet{Vanderburg2016}.}
\label{HD3167_LC}
\end{figure*}

Based on the occurrence rate of planets and planet candidates discovered by \kepler\ we know that short-period super-Earths ($R_\mathrm{p}$\,=\,1\,-\,2\,$R_\oplus$, $M_\mathrm{p}$\,=\,1\,-\,10\,$M_\oplus$) and sub-Neptunes ($R_\mathrm{p}$\,=\,2\,-\,4\,$R_\oplus$, $M_\oplus$\,=\,10\,-\,40\,$M_\oplus$) are extremely common in our Galaxy. About 26\% of solar-like stars in the Milky Way host small planets ($R_\mathrm{p}$\,$<$\,4\,$R_\oplus$) with orbital periods shorter than 100 days \citep[see, e.g.,][]{Marcy2014,Burke2015}. These planets are not represented in our Solar System and were therefore completely unknown to us until a few years ago. 

Although \kepler\ has provided us with a \emph{bonanza} of such small planets, little is known about their masses, compositions, and internal structures. Mass determinations with a precision that allows us to distinguish between different internal compositions (better than 20\,\%) have been possible only for a few dozen super-Earths and sub-Neptunes. The small radial velocity (RV) variation induced by such planets and the faintness of most \kepler\ host stars ($V$\,$>$\,13\,mag) make RV follow-up observations difficult. These observations either place too much demand on telescope time, or they are simply unfeasible with state-of-the-art facilities.

A special class of close-in objects is composed of exoplanets with ultra-short orbital periods \citep[$P_\mathrm{orb}$\,$<$\,1~day; ][]{Sanchis-Ojeda2014}. These planets are the most favorable cases for transit and RV search programs, as the transit probability is high ($\propto$\,$P_\mathrm{orb}^{-2/3}$) and the induced RV variation is large ($\propto$\,$P_\mathrm{orb}^{-1/3}$). Very short orbital periods are also advantageous because they are (often) much shorter than the rotation period of the star, allowing the correlated noise due to stellar rotation to be more easily distinguished from the planet-induced RV signal \citep{Hatzes2011}. To date about 80 ultra-short period exoplanets have been discovered\footnote{See \url{exoplanets.org} and \url{exoplanet.eu}; as of May 2017.}, mainly from transit surveys starting with \object{CoRoT-7b} \citep{Leger2009}. Masses, however, have only been determined for two dozen of these objects. About half of these are gas-giant planets with masses between  1 and 10\,$M_\mathrm{Jup}$. The rest are small planets in the super-earth regime with masses between about 5 and 10~M$_{\oplus}$. 

Using time-series photometric data from the \ktwo\ space mission, \citet{Vanderburg2016} recently announced the discovery of two small transiting planets around the bright (V=8.9~mag) K0 dwarf star \object{HD\,3167}. The inner planet, \object{HD\,3167\,b}, has a radius of $R_\mathrm{p}$=1.6\,$R_\oplus$ and transits the host star every 0.96 days. By our definition, HD\,3167\,b qualifies as an ultra-short period planet. The outer planet, \object{HD\,3167\,c}, has a radius of 2.9\,$R_\oplus$ and an orbital period of 29.85 days. The brightness of the host star makes the system amenable to follow-up observations such as high-precision RV measurements for planetary mass determination.

As part of the ongoing RV follow-up program of \ktwo\ transiting planets successfully carried our by our consortium \texttt{KESPRINT} \citep[e.g.,][]{Sanchis-Ojeda2015,Grziwa2016,VanEylen2016,Barragan2017,Fridlund2017,Guenther2017,Nowak2017}, we herein present the results of an intensive RV campaign we conducted with the FIES, HARPS, and HARPS-N spectrographs to accurately measure the masses of the two small planets transiting HD\,3167. The paper is organized as follows. In \S\,\ref{Sect:lcurve} and \S\,\ref{Sec:RV_FU} we provide a short recap of the \ktwo\ data and describe our high-resolution spectroscopic observations. The properties of the host star are reported in \S\,\ref{Sec:StarProperties}. We present the data modeling in \S\,\ref{Sect:Data_Analysis} along with the frequency analysis of our radial velocity time-series. Results, discussions, and summary are given in \S\,\ref{Sect:Results}~and~\ref{Sect:Discussion}.

\section{\ktwo\ photometry}  
\label{Sect:lcurve}

\ktwo\ observed HD\,3167 during its Campaign~8 for about 80 days -- between January and March 2016 -- with an integration time of about 29.4~minutes (long cadence mode). For our analysis presented in \S\,\ref{Sect:RotationPeriod} and \ref{Sect:JointAnalysis} we used the light curve extracted following the technique described in \citet{Vanderburg2014}\footnote{Publicly available at \url{https://www.cfa.harvard.edu/~avanderb/k2.html}.}. We refer the reader to \citet{Vanderburg2016} for a detailed description of both the \ktwo\ data of HD\,3167 and the detection of the two transiting planets. For the sake of clarity we reproduce in Fig.~\ref{HD3167_LC} the full light curve of HD\,3167 presented in \citet{Vanderburg2016}.

\section{Spectroscopic follow-up}
\label{Sec:RV_FU}

We used the FIbre-fed \'Echelle Spectrograph \citep[FIES;][]{Frandsen1999,Telting2014} mounted at the 2.56m Nordic Optical Telescope (NOT) of Roque de los Muchachos Observatory (La Palma, Spain) to acquire 37 high-resolution spectra (R\,$\approx$\,67000) in 12 different nights between July and September 2016. FIES is mounted inside a heavily insulated building separate from the dome to isolate the spectrograph from sources of thermal and mechanical instability. The temperature inside the building is kept constant within 0.02~$^\circ$C. Observations of RV standard stars performed by our team since 2011, have shown that long-exposed ThAr spectra taken immediately before and after short-exposed target's observations ($T_\mathrm{exp}$\,$\le$20 min) allow us to trace the intra-night RV drift of the instrument to within $\sim$2-3~m/s \citep{Gandolfi2013,Gandolfi2015}, which is comparable with the internal precision of our FIES RV measurements (Table~5). On the other hand, observations of standard stars performed in different nights have shown that the inter-night stability of the instrument is 2 to 4 times worse.

The FIES observations were carried out as part of the OPTICON and NOT observing programs 16A/055, P53-016, and P53-203. We set the exposure time to 15\,-\,20 min and acquired long-exposed ($T_\mathrm{exp}$\,$\approx$\,35 sec) ThAr spectra immediately before and after the target observations. We took at least 2 spectra separated by 1-2 hours per night except on one night. The data were reduced using standard routines, which include bias subtraction, flat fielding, order tracing and extraction, and wavelength calibration. Radial velocities were derived via multi-order cross-correlations, using the stellar spectrum with the highest S/N ratio as a template\footnote{Epoch 2457605.}. The measured RVs are listed in Table~5 along with their 1-$\sigma$ internal uncertainties and the  signal-to-noise (S/N) ratio per pixel at 5500~\AA.

We also acquired 50 spectra with the HARPS spectrograph \citep[R\,$\approx$\,115\,000;][]{Mayor03} and 32 spectra with the HARPS-N spectrograph \citep[R\,$\approx$\,115000;][]{Cosentino2012}. HARPS and HARPS-N are fiber-fed cross-dispersed echelle spectrographs specifically designed to achieve very high-precision long-term RV stabilities ($<$\,1\,\ms). They are mounted at the ESO-3.6m telescope of La Silla observatory (Chile) and at the 3.58m Telescopio Nazionale Galileo (TNG) of Roque de los Muchachos Observatory (La Palma, Spain). 

\begin{table*}[!th]
\begin{center}
\caption{Spectroscopic parameters of HD\,3167 as derived from the FIES (top), HARPS (middle), and HARPS-N (bottom) data using the three methods described in Sect~\ref{Sec:SpecAnalysis}.\label{tab:spec_param}}
\begin{tabular}{lcccccc}
\hline
\hline
\noalign{\smallskip}
Method &  \teff\ (K) & \logg\ (cgs) &  [Fe/H] (dex) &  \vmic\ (\kms) & \vmac\ (\kms) &  \vsini\ (\kms) \\
\hline
\noalign{\smallskip}
\multicolumn{2}{l}{\emph{\bf FIES}} \\
\noalign{\smallskip}
Method 1 & 5288$\pm$75  & 4.53$\pm$0.07 & 0.02$\pm$0.06 & 0.9$\pm$0.1 & 2.3$\pm$0.5 & 1.9$\pm$0.8 \\
Method 2 & 5270$\pm$95  & 4.56$\pm$0.10 & 0.05$\pm$0.05 & 0.9$\pm$0.1 & 2.3$\pm$0.6 & 1.7$\pm$0.6 \\
Method 3 & 5247$\pm$76  & 4.44$\pm$0.19 & 0.01$\pm$0.10 & 0.7$\pm$0.2 & --          &         --  \\
\hline
\noalign{\smallskip}
\multicolumn{2}{l}{\emph{\bf HARPS}} \\
\noalign{\smallskip}
Method 1 & 5295$\pm$70  & 4.54$\pm$0.05 & 0.03$\pm$0.05 & 0.9$\pm$0.1 & 2.4$\pm$0.5 & 1.8$\pm$0.6 \\
Method 2 & 5230$\pm$80  & 4.54$\pm$0.07 & 0.05$\pm$0.06 & 0.9$\pm$0.1 & 2.3$\pm$0.5 & 1.7$\pm$0.6\\
Method 3 & 5257$\pm$112 & 4.41$\pm$0.20 & 0.04$\pm$0.08 & 0.8$\pm$0.1 & -- & -- \\
\hline
\noalign{\smallskip}
\multicolumn{2}{l}{\emph{\bf HARPS-N}} \\
\noalign{\smallskip}
Method 1 & 5275$\pm$62   & 4.51$\pm$0.05 & 0.03$\pm$0.05 & 0.9$\pm$0.1 & 2.4$\pm$0.5 & 1.7$\pm$0.6 \\
Method 2 & 5260$\pm$70   & 4.52$\pm$0.06 & 0.04$\pm$0.05 & 0.9$\pm$0.1 & 2.3$\pm$0.5 & 1.8$\pm$0.6 \\
Method 3 & 5247$\pm$121  & 4.40$\pm$0.20 & 0.06$\pm$0.09 & 0.7$\pm$0.1 & --         &         --   \\
\hline
\end{tabular}
\end{center}
\end{table*}

The HARPS and HARPS-N observations were performed as part of the ESO observing programs 097.C-0948 and 098.C-0860, and of the TNG/CAT programs A33TAC\_15 and CAT16B\_61. We used the simultaneous Fabry Perot calibrator and set the exposure times to 15\,-\,40 minutes depending on sky conditions and scheduling constraints. We followed the same multi-visit strategy adopted for the FIES observations, i.e., we acquired at least 2 spectra per night in most of the observing nights. The data were reduced using the dedicated HARPS and HARPS-N off-line pipelines and radial velocities were extracted by cross-correlating the extracted echelle spectra with a G2 numerical mask. We tested also the K0 and the K5 mask but found neither a significant improvement of the error bars, nor a significant variation of the relative amplitude of the detected RV variation.

The HARPS and HARPS-N radial velocity measurements and their uncertainties are also listed in Table~5, along with the S/N ratio per pixel at 5500~\AA, the full-width half maximum (FWHM) and bisector span (BIS) of the cross-correlation function (CCF), and the Ca\,{\sc ii} H\,\&\,K chromospheric activity index log\,$R^\prime_\mathrm{HK}$. Five out of the 50 HARPS spectra are affected by poor sky and seeing conditions. They are not listed in Table~5 and were not used in our analysis.

\section{Stellar properties}
\label{Sec:StarProperties}

\subsection{Spectroscopic parameters}
\label{Sec:SpecAnalysis}

We combined separately the FIES, HARPS, and HARPS-N data to produce three co-added spectra of higher S/N ratio and determine the spectroscopic parameters of the host star. The stacked FIES, HARPS, and HARPS-N spectra have S/N ratios of 500, 560, and 480 per pixel at 5500~\AA, respectively. We derived the spectroscopic parameters using three independent methods as described in the next three paragraphs. Results for each method and spectrum are listed in Table~\ref{tab:spec_param}.

\emph{-- Method 1}. This uses a customized \texttt{IDL} software suite that implements the spectral synthesis program \texttt{SPECTRUM}\footnote{Publicly available at \url{http://www.appstate.edu/~grayro/spectrum/spectrum.html}.} \citep[V2.76;][]{Gray1994} to compute synthetic spectra using \texttt{ATLAS\,9} model atmospheres \citep{Castelli2004}. The code fits spectral features that are sensitive to different photospheric parameters, adopting the calibration equations of \citet{Bruntt2010} and \citet{Doyle2014} to determine the microturbulent (\vmic) and macroturbulent (\vmac) velocities. It uses the wings of the Balmer lines to obtain a first guess of the effective temperature (\teff), and the Mg\,{\sc i}~5167, 5173, 5184~\AA, the Ca\,{\sc i}~6162, 6439~\AA, and the Na\,{\sc i}~D lines to refine the effective temperature estimate and derive the surface gravity (\logg). The iron abundance [Fe/H] and projected rotational velocity \vsini\ are measured by fitting many isolated and unblended iron lines.

\emph{-- Method 2}. This uses the spectral analysis package \texttt{SME} \citep[V4.43;][]{Valenti1996,Valenti2005} along with both \texttt{ATLAS\,12} and \texttt{MARCS} model atmospheres \citep{Kurucz2013,Gustafsson2008}. \texttt{SME} calculates synthetic spectra and fits them iteratively to the observed high-resolution echelle spectra using a chi-squared minimization procedure. Micro and macro turbulent velocities are estimated using the same calibration equations adopted by the first method. \teff, \logg, [Fe/H], and \vsini\ are derived by fitting the same spectral features as in the previous paragraph.

\emph{-- Method 3}. This is based on the classical equivalent width (EW) technique applied to about 100 Fe\,{\sc i} and 10 Fe\,{\sc ii} lines. It uses the public version of the line list prepared for the {\it Gaia}-ESO Survey \citep{Heiter2015}, which is based on the \texttt{VALD3} atomic database \citep{Ryabchikova2011}.  \teff\ is obtained by removing trends between the abundance of a given element and the respective excitation potential; \logg\ is derived by assuming the ionization equilibrium condition, i.e., by requiring that for a given species the same abundance (within the uncertainties) is obtained from lines of two ionization states (typically neutral and singly ionized species); \vmic\ and [Fe/H] are estimated by minimizing the slope of the relationship between abundance and the logarithm of the reduced EWs. Equivalent widths are measured using the code \texttt{DOOp} \citep{Cantat2014}, a wrapper of \texttt{DAOSPEC} \citep{Stetson2008}. The photospheric parameters are derived with the code \texttt{FAMA} \citep{Magrini2013}, a wrapper of \texttt{MOOG} \citep{Sneden2012}.

The three techniques provide consistent results, regardless of the used spectrum and/or method. While we have no reason to prefer one method over the other, we adopted the results of \emph{Method 1} applied on the FIES, HARPS, and HARPS-N spectra. The final adopted values for \teff, \logg, [Fe/H], and \vsini\ are the averaged estimates we obtained using the first method; the corresponding uncertainties are defined as the square root of the individual errors added in quadrature, divided by three. We obtained \teff\,=\,5286$\pm$40~K, \logg\,=\,4.53$\pm$0.03~(cgs), [Fe/H]\,=\,0.03$\pm$0.03~dex, and \vsini\,=\,1.8$\pm$0.4~\kms\ (Table~\ref{Tab:StellarParameter}). Our results are in fairly good agreement with the spectroscopic parameters derived by \citet{Vanderburg2016}.

\subsection{Stellar mass, radius, and age}
\label{Mass_radius_age}

We followed the same method adopted by \citet{Vanderburg2016} and derived the mass, radius, and age of HD\,3167 using \texttt{PARAM}, an on-line interface for Bayesian estimation of stellar parameters\footnote{Available at \url{http://stev.oapd.inaf.it/cgi-bin/param}.}. Briefly, \texttt{PARAM} interpolates the apparent visual magnitude, parallax, effective temperature and iron abundance onto \texttt{PARSEC} model isochrones \citep{Bressan2012}, adopting the initial mass function from \citet{Chabrier2001}. We used our spectroscopic parameters (\S\,\ref{Sec:SpecAnalysis}) along with the V-band magnitude listed in the EPIC input catalog ($V$\,=\,8.941\,$\pm$\,0.015~mag) and the \texttt{Hipparcos}' parallax \citep[21.82\,$\pm$\,1.05~mas,][]{vanLeeuwen2007}\footnote{{\it Gaia}'s first data release does not report the parallax of HD\,3167.}. Following the method outlined in \citet{Gandolfi2008} and using the broad-band photometry available in the EPIC input catalog, we found that the interstellar reddening is indeed consistent with zero ($A_\mathrm{v}$\,=\,0.02\,$\pm$\,0.03~mag), as expected given the short distance to the star (45.8\,$\pm$\,2.2~pc). We set the interstellar absorption to zero and did not correct the apparent visual magnitude. 

HD\,3167 has a mass of $M_\star$\,=\,\smass\ and a radius $R_\star$\,=\,\sradius, implying a surface gravity of \logg\,=\,4.51\,$\pm$\,0.03 (cgs), in agreement with the spectroscopically derived value (\S\,\ref{Sec:SpecAnalysis}). The isochrones constrain the age of the star to be 5\,$\pm$\,4~Gyr.

\subsection{Stellar activity and rotation period}
\label{Sect:RotationPeriod}

The average Ca\,{\sc ii} H\,\&\,K activity index log\,$R^\prime_\mathrm{HK}$, as measured from the HARPS and HARPS-N spectra, is\,\,$-$5.03\,$\pm$\,0.01 and $-$5.06\,$\pm$\,0.02 dex, respectively, indicative of a low chromospheric activity level\footnote{As comparison, the activity index of the Sun varies between $-5.0$ and $-4.8$~dex.}. We checked if the extrinsic absorption, either from the interstellar medium (ISM) or from material local to the system, biases the measured values of log\,$R^\prime_\mathrm{HK}$ \citep{Fossati2013,Fossati2015}. The far-ultraviolet (FUV) stellar emission, which originates in the chromosphere and transition region, provides instead an unbiased measure of the stellar activity \citep{Fossati2015}. We measured the excess of the chromospheric FUV emission -- directly proportional to stellar activity -- by estimating the difference between the measured GALEX FUV flux and the photospheric flux derived from a MARCS model with the same photospheric parameters as the star \citep{Gustafsson2008} rescaled to fit the observed optical (Johnson and Tycho) and infrared (2MASS and WISE) photometry of HD\,3167. The fit accounts for the interstellar extinction reported in \S\,\ref{Mass_radius_age}. We obtained an excess emission in the GALEX FUV band of about 260~erg\,cm$^{-2}$\,s$^{-1}$, indicative of a low level of stellar activity \citep{Shkolnik2014}, in agreement with the log\,$R^\prime_\mathrm{HK}$ value. This provides evidence that the Ca\,{\sc ii} activity index log\,$R^\prime_\mathrm{HK}$ is very likely not biased by extrinsic absorption.

The light curve of HD\,3167 displays a 0.08\,\% flux drop occurring during the first half of the \ktwo\ observations and lasting for about 35-40 days (Fig.~\ref{HD3167_LC}). If the variation were due to an active region moving in and out of sight as the star rotates around its axis, then the rotation period of the star should be at least twice as long, i.e., 70-80 days. Such a long rotation period seems to be unlikely for a K-type dwarf and is inconsistent with our \vsini\ measurement and stellar radius determination (see below). Figure~\ref{Period_Histogram} shows the distribution of the rotation period of 3591 late G- and early K-type dwarfs as measured by \citet{McQuillan2014} using \kepler\ light curves. We selected only \kepler\ stars with photospheric parameters similar to those of HD\,3167, i.e., objects with 5170\,$\le$\,\teff\,$\le$\,5370\,K and \logg\,$\ge$\,4.40 (cgs). None of the ``HD\,3167's \kepler\ twins'' has a rotation period longer than 70 days. Moreover, only 9 objects have a rotational period exceeding 50 days. As both the K2 light curves of HD\,3167 -- as extracted by \citet{Luger2016} and \citet{Aigrain2016} -- display the same feature, we conclude that the observed 0.08\,\% flux drop is very likely an instrumental artifact caused by the spacecraft pointing jitter.

\begin{figure}
\includegraphics[width=\columnwidth]{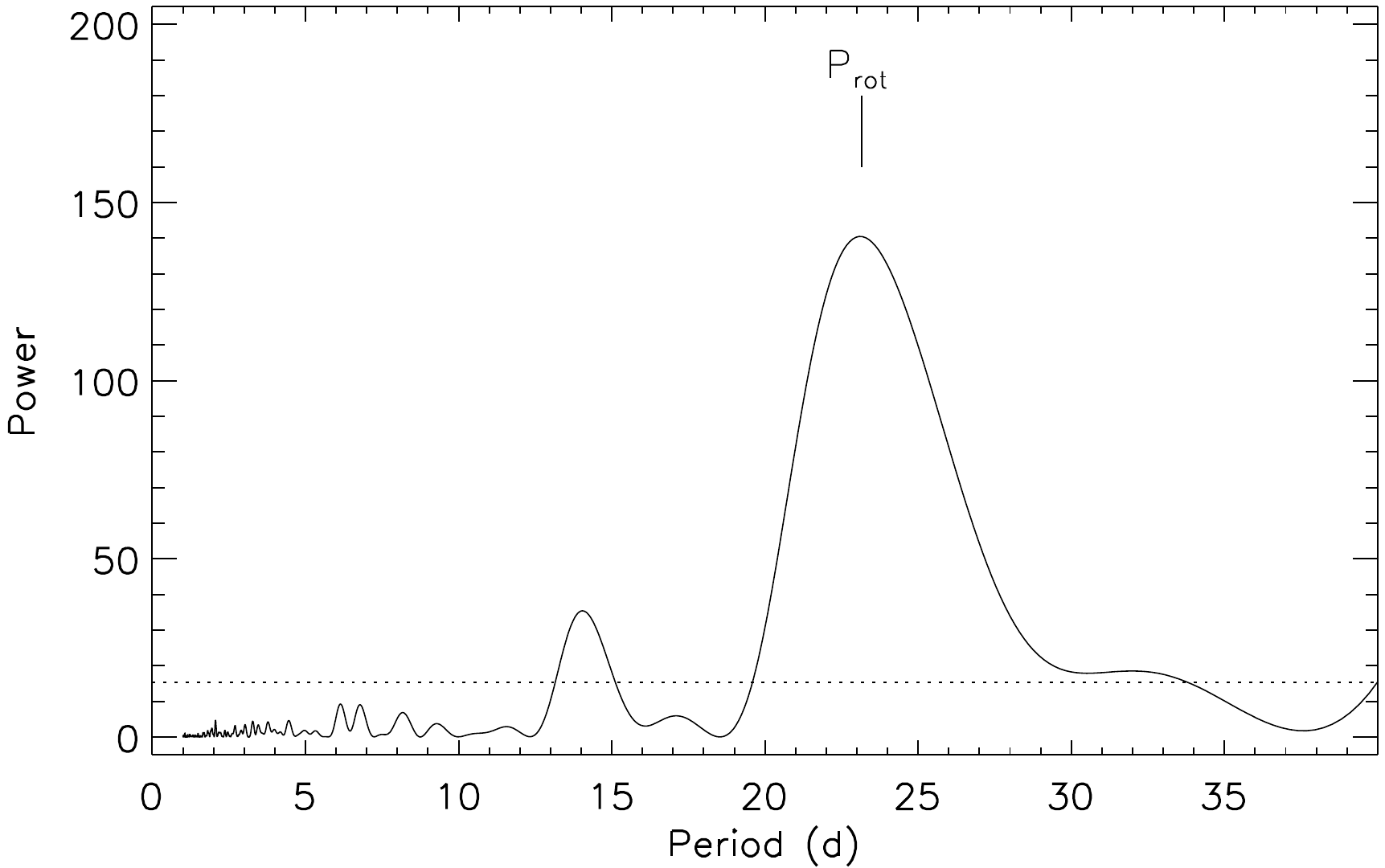}
\caption{Lomb-Scargle periodogram of the \ktwo\ light curve of HD\,3167. The horizontal dashed line marks the 0.1\% FAP as defined in \citet{Scargle1982}.}
\label{HD3167_LC_Periodogram}
\end{figure}

Figure~\ref{HD3167_LC_Periodogram} shows the Lomb-Scargle (LS) periodogram \citep{Lomb1976, Scargle1982} of the \ktwo\ light curve of HD\,3167 following the subtraction of the best fitting transit models of planet b and c (\S\,\ref{Sect:Data_Analysis}). Besides a very strong peak at $\sim$75 days due to the flux drop described in the previous paragraph, there are 2 additional significant peaks at 14 and 23.5 days with a Scargle's false alarm probability (FAP) lower than 0.1\,\%. Since the period ratio is close to 0.5, we interpreted the former as the first harmonic of the latter. With an amplitude of about 0.04\,\%, the 23.5-day signal is clearly visible in the first half of the \ktwo\ time series data, whereas is barely visible in the second half of the photometric data (Fig.~\ref{HD3167_LC}). As a sanity check, we split the light curve into two chunks of $\sim$40 days and calculated the LS periodogram of each chunk. The 23.5-day signal is detected also in the second half of the light curve but with a lower significance. This is likely due to the 80\,\% higher noise level of the second half of the \ktwo\ data with respect to the first half, as pointed out by \citet{Vanderburg2016}.

\begin{figure}
\includegraphics[width=\columnwidth]{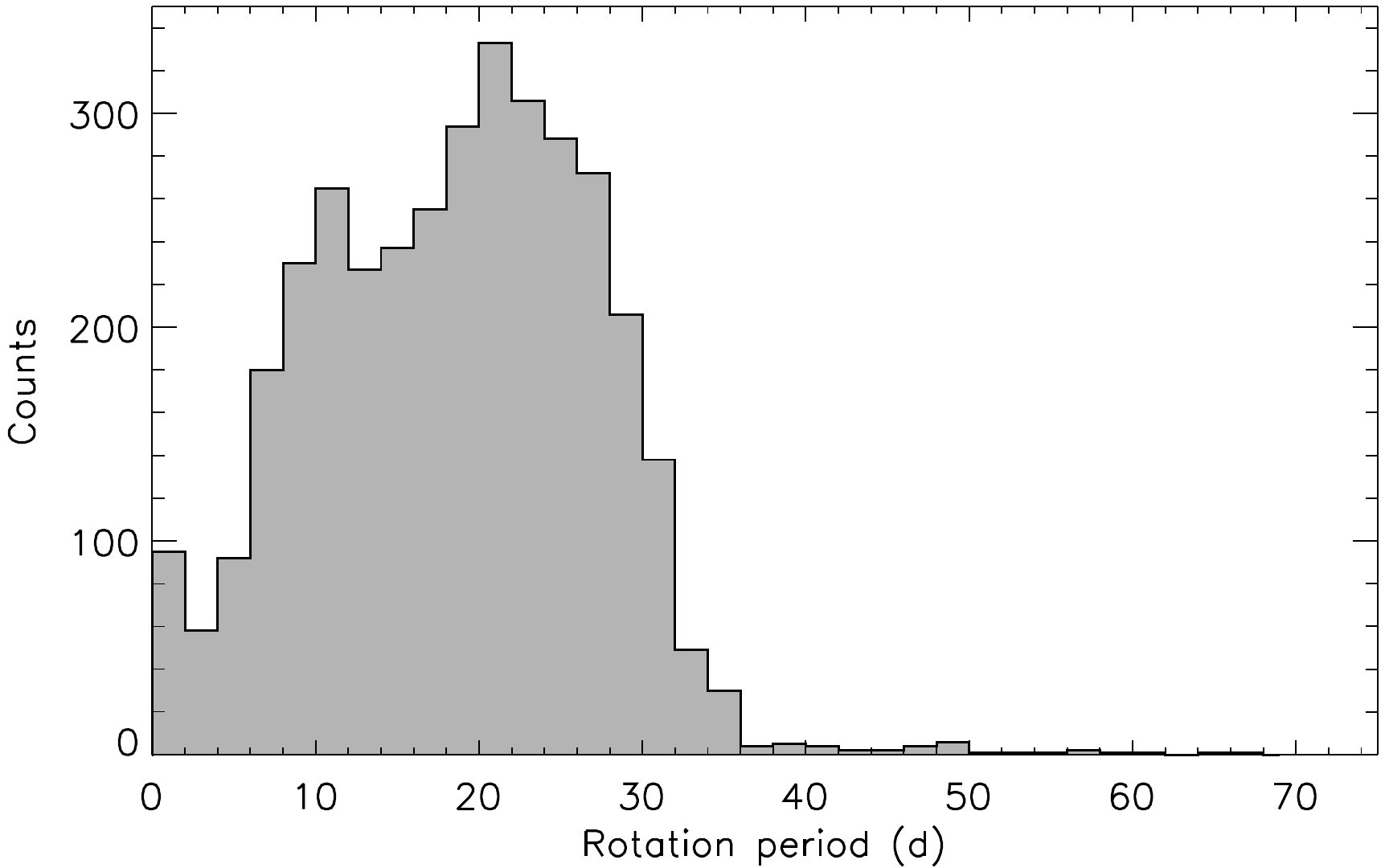}
\caption{Rotation period distribution of \kepler\ field stars with 5170\,$\le$\,\teff\,$\le$\,5370\,K and \logg\,$\ge$\,4.3 (cgs), as extracted from the work of \citet{McQuillan2014}.}
\label{Period_Histogram}
\end{figure}

We interpreted the 23.5-day signal as the rotation period of the star and attributed the peak at 14 days to the presence of active regions located at opposite stellar longitudes. We measured a rotation period and uncertainty of $P_\mathrm{rot}$\,=\,23.52\,$\pm$\,2.87 days defined as the position and full width at half maximum of the corresponding peak in the LS periodogram. If the rotation period of the star were instead 14 days, the magnetic activity of the star would very likely be stronger than what measured from the log\,$R^\prime_\mathrm{HK}$ activity index \citep{Suarez2015}. It is also worth noting that the distribution of the rotational periods of HD\,3167's \kepler\ twins is peaked between 20 and 25 days (Fig.~\ref{Period_Histogram}).

The spectroscopically derived projected rotational velocity of the star \vsini\,=\,1.8\,$\pm$\,0.4\,\kms, combined with the stellar radius $R_\star$\,=\sradius, implies an upper limit on the rotation period of 23.5\,$\pm$\,5.3 days, in agreement with the period derived from the K2 light curve, further corroborating our results. This also suggests that the star is seen nearly equator-on and that the transiting multi-planet system around HD\,3167 might be aligned along the line-of-sight.

\begin{table}[t]
\begin{center}
\caption{Stellar parameters.}
\label{Tab:StellarParameter}
\begin{tabular}{lc}
\hline
\hline
\noalign{\smallskip}
Parameter & Value  \\
\noalign{\smallskip}
\hline
\noalign{\smallskip}
Effective Temperature \teff\ (K)             & 5286\,$\pm$\,40   \\
Surface gravity$^{(\mathrm{a})}$ \logg (cgs) & 4.53\,$\pm$\,0.03 \\
Surface gravity$^{(\mathrm{b})}$ \logg (cgs) & 4.51\,$\pm$\,0.03 \\
Iron abundance [Fe/H] (dex)                  & 0.03\,$\pm$\,0.03 \\
Projected rot. velocity \vsini (\kms)        & 1.8\,$\pm$\,0.4  \\
Interstellar extinction $A_\mathrm{v}$ (mag) & 0.02\,$\pm$\,0.03 \\
Stellar mass $M_\star$ ($M_\odot$)           & \smass[]    \\
Stellar mass $R_\star$ ($M_\odot$)           & \sradius[]  \\
Age (Gyr)                                    & 5\,$\pm$\,4 \\
Rotation period $P_\mathrm{rot}$ (day)       & 23.52\,$\pm$\,2.87 \\
Distance$^{(\mathrm{c})}$ (pc)               & 45.8\,$\pm$\,2.2 \\
\noalign{\smallskip}
\hline
\end{tabular}
\end{center}
\tablenotetext{a}{From spectroscopy.}
\tablenotetext{b}{From spectroscopy and isochrones.}
\tablenotetext{c}{\texttt{Hipparcos}' distance from \citet{vanLeeuwen2007}.}
\end{table}

\section{Data analysis}
\label{Sect:Data_Analysis}

\subsection{Periodogram analysis of the radial velocities}
\label{Sect:GLS_RVs}

We first performed a frequency analysis of the RV measurements in order to look for possible periodic signals in the data and assess if, in the absence of the K2 transit photometry, we would have been able to detect the presence of HD\,3167\,b and c. For this purpose we used only the HARPS and HARPS-N measurements because of the higher quality and superb long-term stability of the two instruments.

\begin{figure}
\includegraphics[width=\columnwidth]{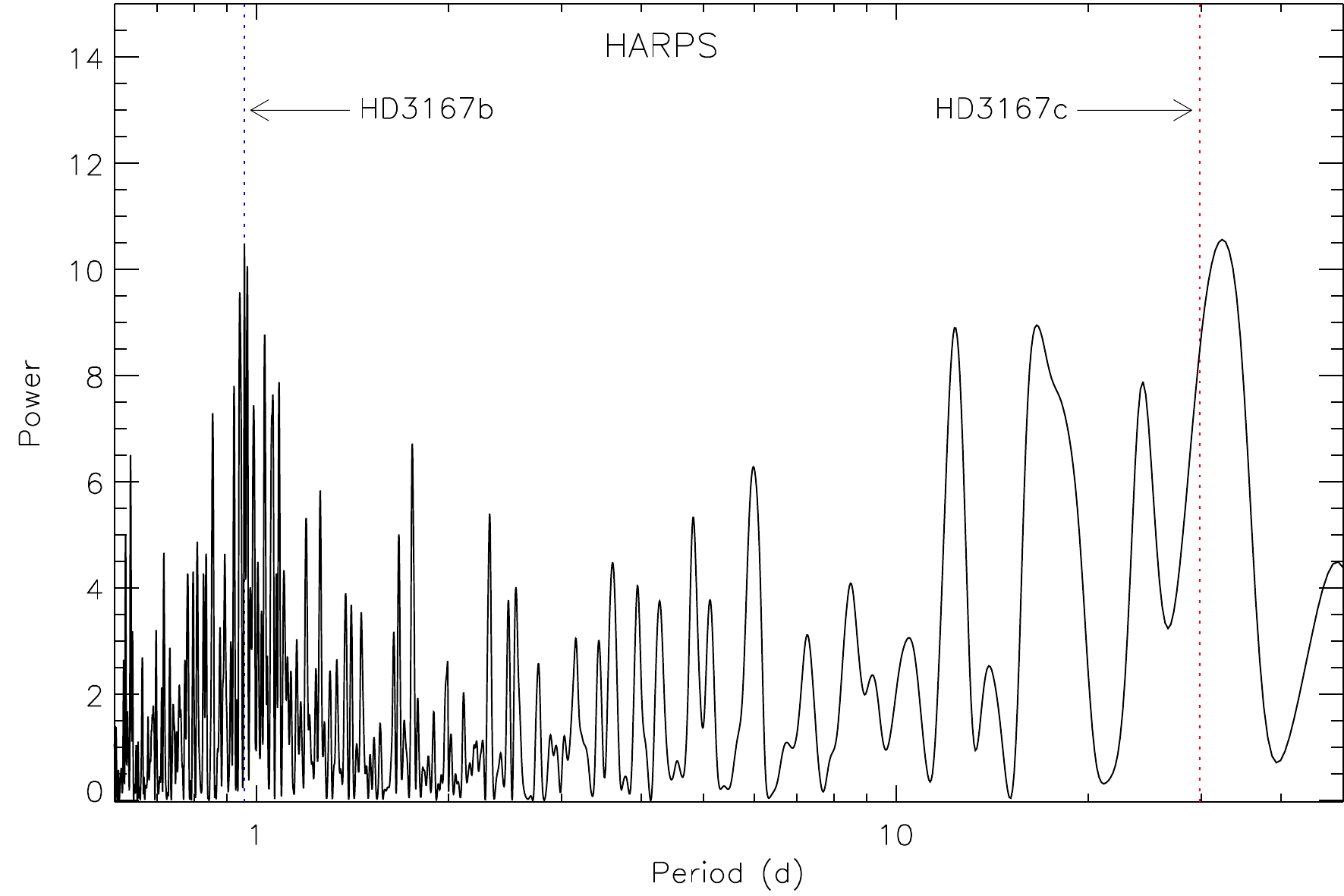}
\includegraphics[width=\columnwidth]{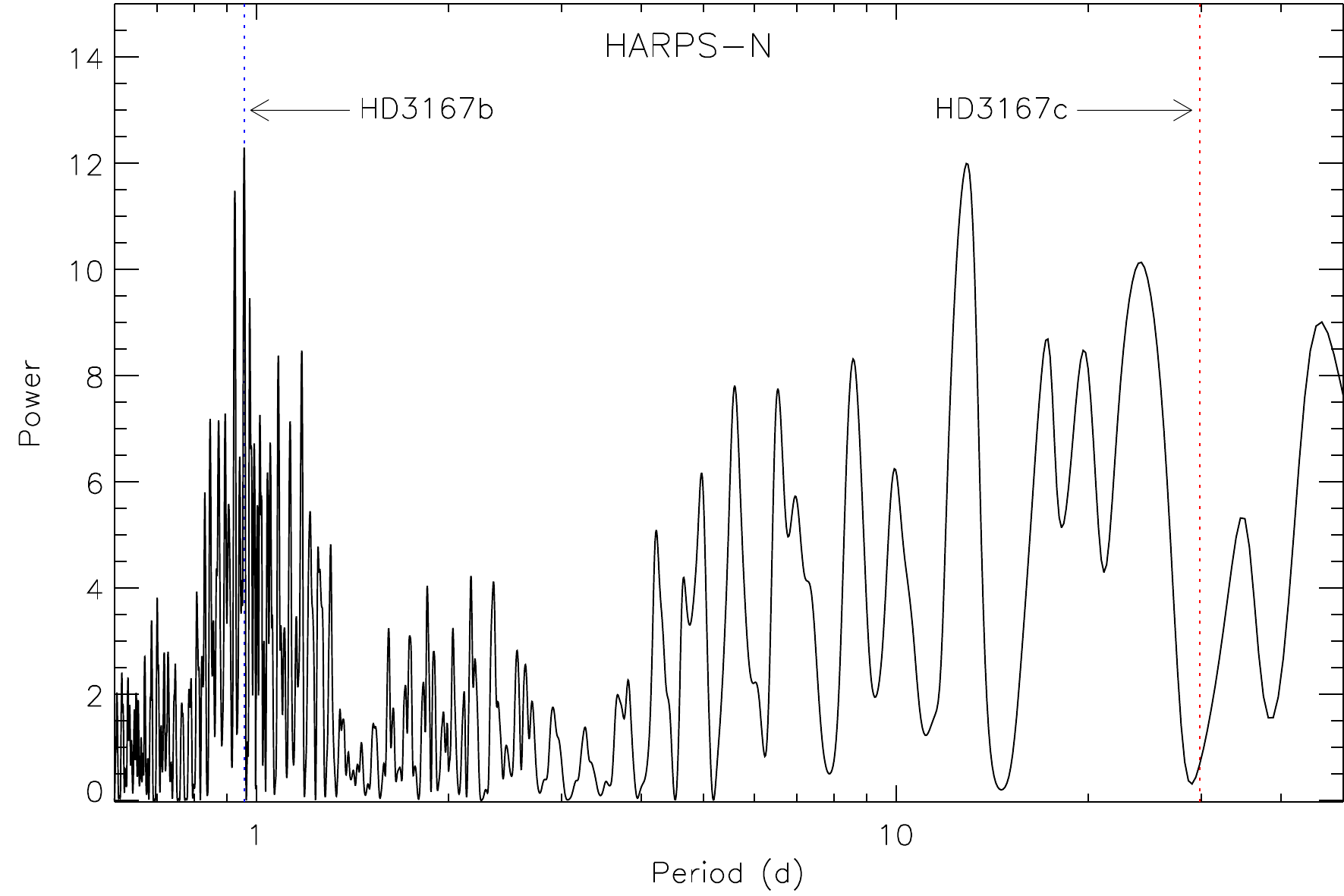}
\includegraphics[width=\columnwidth]{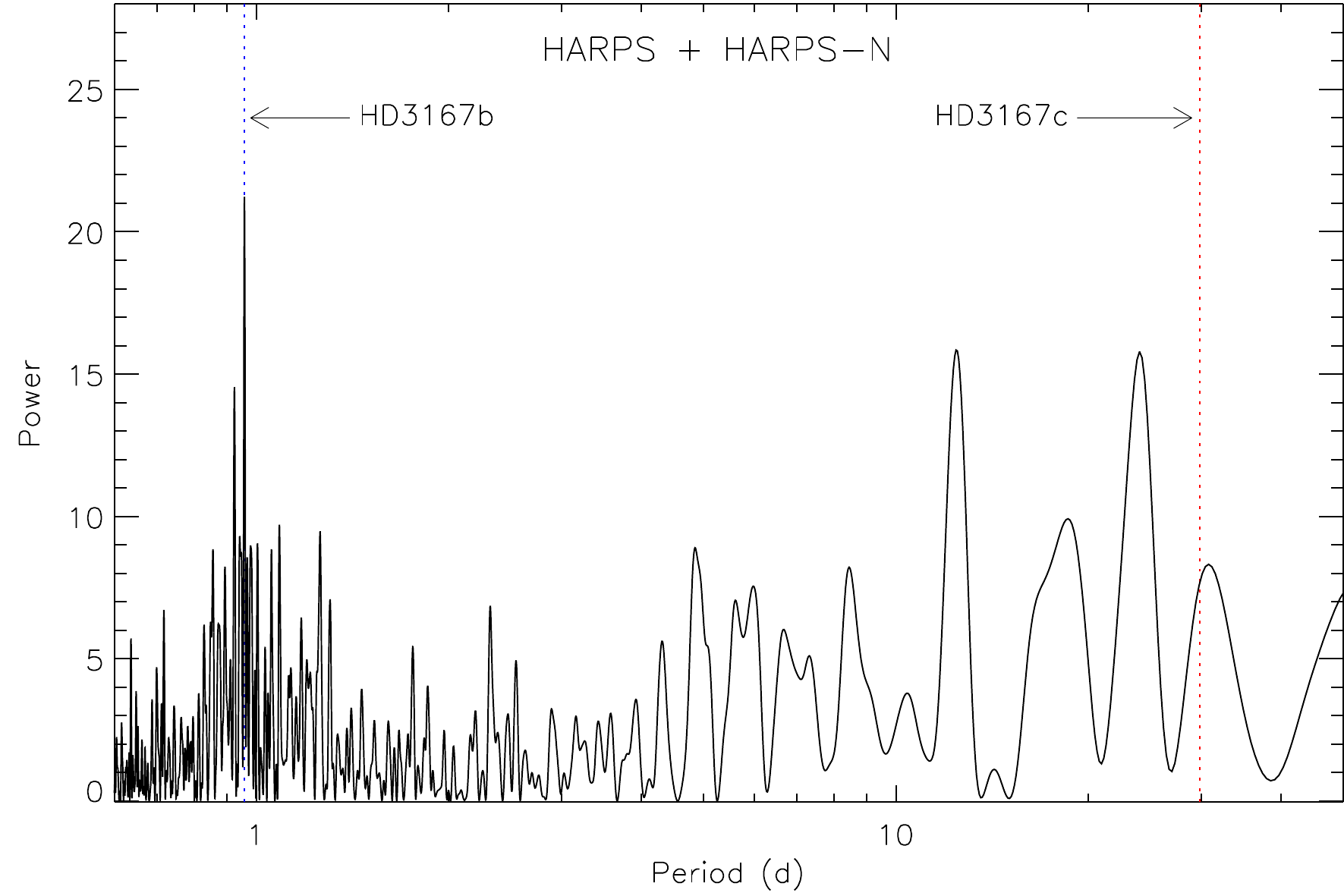}
\caption{GLS periodograms of the HARPS (top panel), HARPS-N (middle panel), and HARPS+HARPS-N (bottom panel) RV measurements. The vertical dashed lines mark the orbital periods of HD\,3167\,b (0.96~d) and HD\,3167\,c  (29.85~d).}
\label{periodogram}
\end{figure}

We first analyzed the two data sets separately to account for the velocity offset between the two spectrographs. Although HARPS and HARPS-N are very similar, a small offset ($<$10\,\ms) is expected given, e.g., the different detector, optics, wavelength coverage of the two instruments. The generalized Lomb-Scargle \citep[GLS;][]{Zechmeister2009} periodograms of the HARPS and HARPS-N RVs show a significant peak at the orbital period of HD\,3167\,b, with a false alarm probability\footnote{The FAPs reported in this subsection have been calculated using Eq.~24 of \citet{Zechmeister2009} and should be regarded as preliminary estimates. Deriving reliable FAPs through a bootstrap analysis -- as presented in \S\,\ref{Sect:ResidualAnalysis} -- goes beyond the scope of this subsection.} (FAP) of about 10$^{-5}$ and 10$^{-7}$, respectively (top and middle panel of Figure~\ref{periodogram}). We conclude that the signal of the inner planet HD\,3167\,b is clearly present in both data sets. The GLS periodogram of the HARPS data displays a significant peak at $\sim$32 days (FAP=10$^{-4}$), which is close to the orbital period of HD\,3167\,c (29.85 days). However, the outer transiting planet remains undetected in the HARPS-N data, owing to the uneven sampling of the orbital phase of the outer transiting planet with this instrument (Fig.~\ref{Fig:Joint_Planet_b_c}).

On three occasions\footnote{Epochs 2457611, 2457646, and 2457692.} we observed HD\,3167 nearly simultaneously (within 10 minutes) with HARPS and HARPS-N. We used these measurements to measure the offsets of the RV, FWHM, BIS and log\,$R^\prime_\mathrm{HK}$ between the two data sets and calculate the periodograms of the combined data. We found $\Delta$\,RV$_\mathrm{(HS-HN)}$\,=\,8.0\,$\pm$\,0.5\,\ms, $\Delta$\,FWHM$_\mathrm{(HS-HN)}$\,=\,0.068$\pm$0.006\,\kms, $\Delta$\,BIS$_\mathrm{(HS-HN)}$ = 0.009\,$\pm$\,0.003\,\kms, and $\Delta$\,log\,$R^\prime_\mathrm{HK}$\,$_\mathrm{(HS-HN)}=-0.030 \pm 0.005$. We stress that these offsets have only been used to perform the periodogram analysis of the joint data.

As expected, the GLS periodogram of the joint data set (bottom panel of Fig.~\ref{periodogram}) shows a very significant peak at the orbital period of the inner planet HD\,3167\,b (FAP=10$^{-10}$) and a moderately significant peak at the orbital period of HD\,3167\,c. 

It's worth noting that the three periodograms show also the presence of a significant peak at 23.8~days (0.042~c/d), which is close to the rotation period of the star. We stress, however, that this peak corresponds to the 1-day alias of the orbital period of HD\,3167\,b. The periodogram of the RV residuals -- as obtained following the subtraction of the signals of the two planets -- show no peaks at 0.042~c/d (Fig.~\ref{GLS_RV_Residuals}). 

\subsection{Orbital solution of HD\,3167\,b}
\label{HD3167b_FCO}

We performed a Keplerian fit of the FIES, HARPS, and HARPS-N RV data following the floating chunk offset (FCO) method described in \citet{Hatzes2011}. The FCO method exploits the reasonable assumption that, for ultra-short period planets, RV measurements taken on a single night mainly reflect the orbital motion of the planet rather than other, longer period phenomena such as stellar rotation, magnetic activity, and additional planets. If we can sample a sufficient time segment of the Keplerian curve, then these nightly ``chunks'' can be shifted until the best fit to the orbital motion is found. This method was successfully used to determine the mass of the ultra-short period planets CoRoT-7b \citep{Hatzes2011} and \object{Kepler-78b} \citep{Hatzes2014}.

The ultra-short period planet HD\,3167\,b is well suited for application of the FCO method. This technique is particularly effective at filtering out the RV jitter due to activity. The star has an estimated rotation period of about 23.5\,days (\S\,\ref{Sect:RotationPeriod}), which is longer than the orbital period of HD\,3167\,b. Although HD\,3167 is a relatively inactive star, the FCO method helps in filtering out even a small amount of activity. HD\,3167\,c has an orbital period of about 29.95 days, which results in a change of less than 0.01 in phase within the nightly visibility window of the target ($\sim$5-6 hours). The RV of the star due to the outer transiting planet does not change significantly during an observing night. Moreover, each of the three data sets has its own zero-point offset, which is naturally taken into account by the method. Finally, the FCO technique also removes -- or at least greatly minimizes -- any long term systematic errors, such as the night-to-night RV drifts of FIES (\S\,\ref{Sec:RV_FU}).

We modeled the FIES, HARPS and HARPS-N RV measurements with our code \texttt{pyaneti} \citep{Barragan2016}, a MCMC-based software suite that explores the parameter space using the ensemble sampler with the affine invariance algorithm \citep{Goodman2010}. Following \citet{Hatzes2011}, we divided the RVs into three subsets of nightly measurements -- one per instrument -- and analyzed only those radial velocities for which multiple measurements were acquired on the same night, leading to a total of 12, 15, and 11 chunks of nightly FIES, HARPS, and HARPS-N RVs, respectively. The best fitting orbital solution of HD\,3167\,b was found keeping period and transit ephemeris fixed to the values derived by our joint analysis described in \S\,\ref{Sect:JointAnalysis}, but allowing the RV semi-amplitude variation $K_\mathrm{b}$ and the 38 nightly offsets to vary. We also fitted for $\sqrt{e_\mathrm{b}}\,\sin \omega_\mathrm{\star, b}$ and $\sqrt{e_\mathrm{b}}\,\cos \omega_\mathrm{\star, b}$, where $e_\mathrm{b}$ is the eccentricity and $\omega_\mathrm{\star, b}$ is the argument of periastron of the star \citep{Ford2006}. We adopted uniform uninformative priors within a wide range for each parameter and ran 500 independent Markov chains. The burn-in phase was performed with 25000 iterations using a thin factor of 50, leading to a posterior distribution of 250000 independent data points for each fitted parameter. The final estimates and their 1-$\sigma$ uncertainties were taken as the median and the 68\,\% of the credible interval of the posterior distributions.

\begin{figure}
\includegraphics[width=\columnwidth]{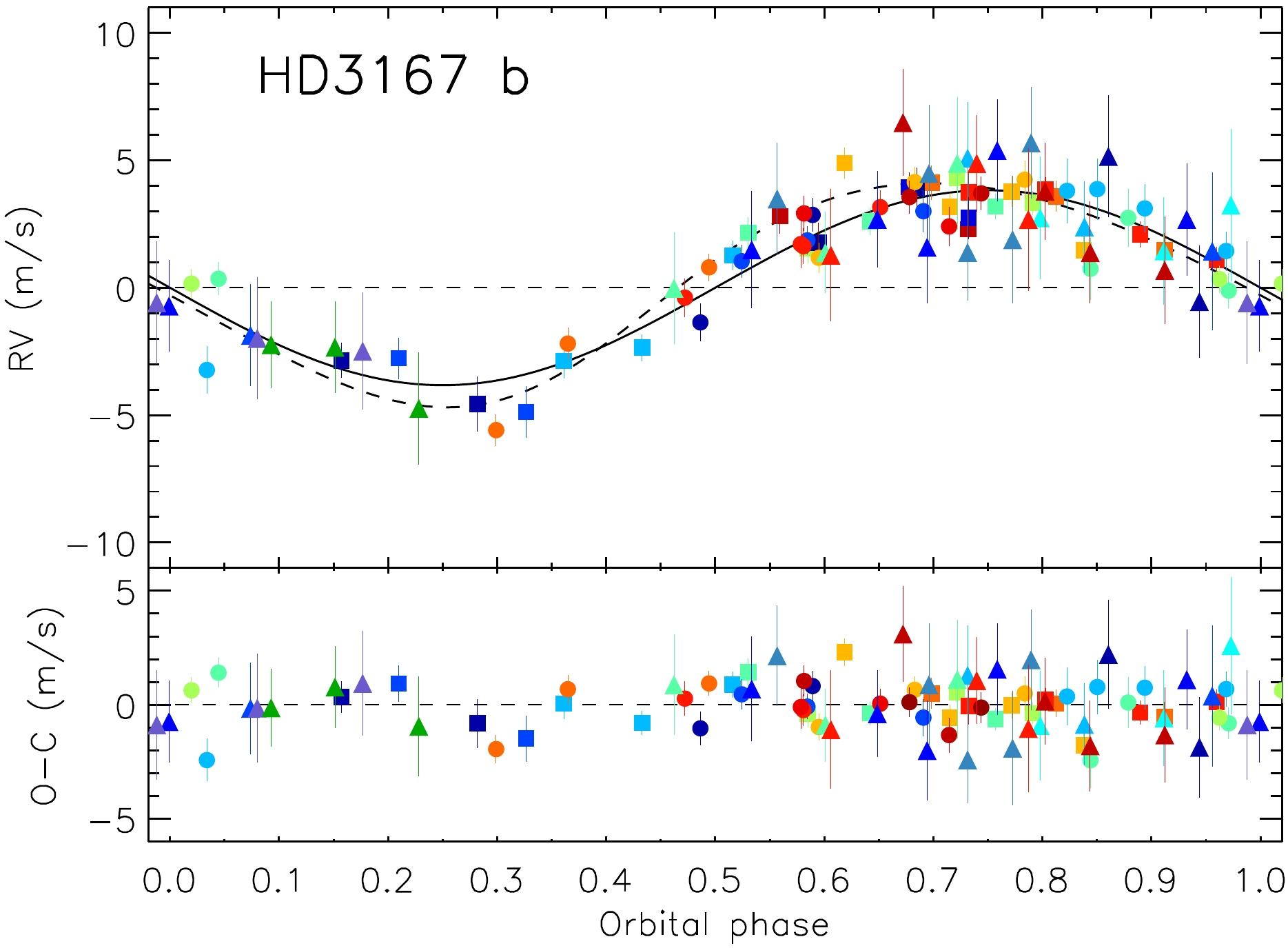}
\caption{\emph{Upper panel}: FIES (triangles), HARPS (circles), and HARPS-N (squares) RV measurements and circular orbital solution of HD\,3167\,b (solid line) derived using the FCO method. Different colors represent measurements for different observing nights. \emph{Lower panel}: Residuals to the circular model.
\label{fig:hd3167bfco}}
\end{figure}

We obtained a best fitting non-zero eccentricity of $e_\mathrm{b}$\,=\,0.112\,$\pm$0.024. We also fitted the RV data assuming a circular orbit ($\sqrt{e_\mathrm{b}}\,\sin \omega_\mathrm{\star, b}$\,=\,$\sqrt{e_\mathrm{b}}\,\cos \omega_\mathrm{\star, b}$\,=\,0). Figure~\ref{fig:hd3167bfco} displays our FIES, HARPS, and HARPS-N measurements along with the best fitting circular (thick line) and eccentric model (dashed line). Different symbols refers to different instrument, whereas different colors represent different nights. We note that the best fitting eccentric solution might be driven by the uneven distribution of data points along the RV curve (Fig.~\ref{fig:hd3167bfco}). In order to asses the significance of our result we created $10^5$ sets of synthetic RVs that sample the best fitting circular solution at the epochs of our real observations. We added Gaussian noise at the same level of our measurements and fitted the simulated data allowing for an eccentric solution. We found that there is a $\sim$7\,\% probability that the best fitting eccentric solution could have arisen by chance if the orbit were actually circular. As this is above the 5\,\% significance level suggested by \citet{Lucy1971}, we decided to conservatively assume a circular model. We found a radial velocity semi-amplitude variation of $K_\mathrm{b}$\,=\,3.82\,$\pm$\,0.42~\ms\, which translates into a mass of $M_\mathrm{b}$\,=\,5.40\,$\pm$\,0.60\,$M_\oplus$ for HD\,3167\,b. We note that the eccentric solution provides a planetary mass that is consistent within 1-$\sigma$ with the result from the circular model.

\subsection{Transit and RV joint analysis}
\label{Sect:JointAnalysis}

We performed a joint modeling of the \ktwo\ and RV measurements with \texttt{pyaneti}. The photometric data includes 6 and 15 hours of \ktwo\ data-points centered around each transit of HD\,3167\,b~and~c. We detrended the segments using the program \texttt{exotrending}\footnote{Available at \url{https://github.com/oscaribv/exotrending}.}. Briefly, we fitted a second order polynomial to the out-of-transit data and removed outliers using a 3-sigma-clipping algorithm applied to the residuals of the preliminary best fitting transit models derived using the formalism of \citet{mandel_2002} coupled to a non-linear least square fitting procedure. As for the RV data sets, we used only the HARPS and HARPS-N measurements because of the long-term instability of the FIES spectrograph (\S\,\ref{Sec:RV_FU}).

We modeled the RV data with two Keplerian signals and fitted the transit light curves using the \citet{mandel_2002}'s model with a quadratic limb darkening law. To account for the \ktwo\ long cadence data, we integrated the transit models over 10 steps. We adopted the same Gaussian likelihood as defined in \citet{Barragan2016}. For each planet $i$ we fitted for the orbital period $P_i$, time of first transit $T_{0,i}$, scaled semi-major axis $a_i/R_{\star}$, impact parameter $b_i$,  planet-to-star radius ratio $R_{\rm i}/R_{\star}$, and RV semi-amplitude variation $K_i$. We assumed a circular orbit for the inner planet and fitted for $\sqrt{e_c} \sin \omega_{\star, c}$ and $\sqrt{e_c} \cos \omega_{\star, c}$ for the outer planet. 

The 30-minute integration time of \ktwo\ smears out the shape of planetary transits increasing the degeneracy between the scaled semi-major axis $a/R_{\star}$ and the impact parameter $b$ \citep{Csizmadia2011}. We therefore set Gaussian priors for the stellar mass and radius using the values derived in \S\,\ref{Mass_radius_age} and constrained $a_\mathrm{i}/R_\star$ of both planets from their orbital periods through Kepler's third law.

We explored the parameter space with 500 independent chains created from random priors for each parameter, as listed in the second column of Table~\ref{Tab:Parameters}. The convergence of the MCMC chains was checked with the Gelman-Rubin statistic. Once all chains converged, we ran 25000 more iterations with a thin factor of 50. This led to a posterior distribution of 250000 independent points for each fitted parameter.

The two-planet model provides a poor fit to the HARPS and HARPS-N measurements with a RV $\chi^2$ of 597 and $\chi^2/{\rm dof} = 8.7$, suggesting that additional signals might be present in the data, as discussed in the next section. 

\subsection{Frequency analysis of the RV residuals}
\label{Sect:ResidualAnalysis}

\begin{figure}
\includegraphics[width=\columnwidth]{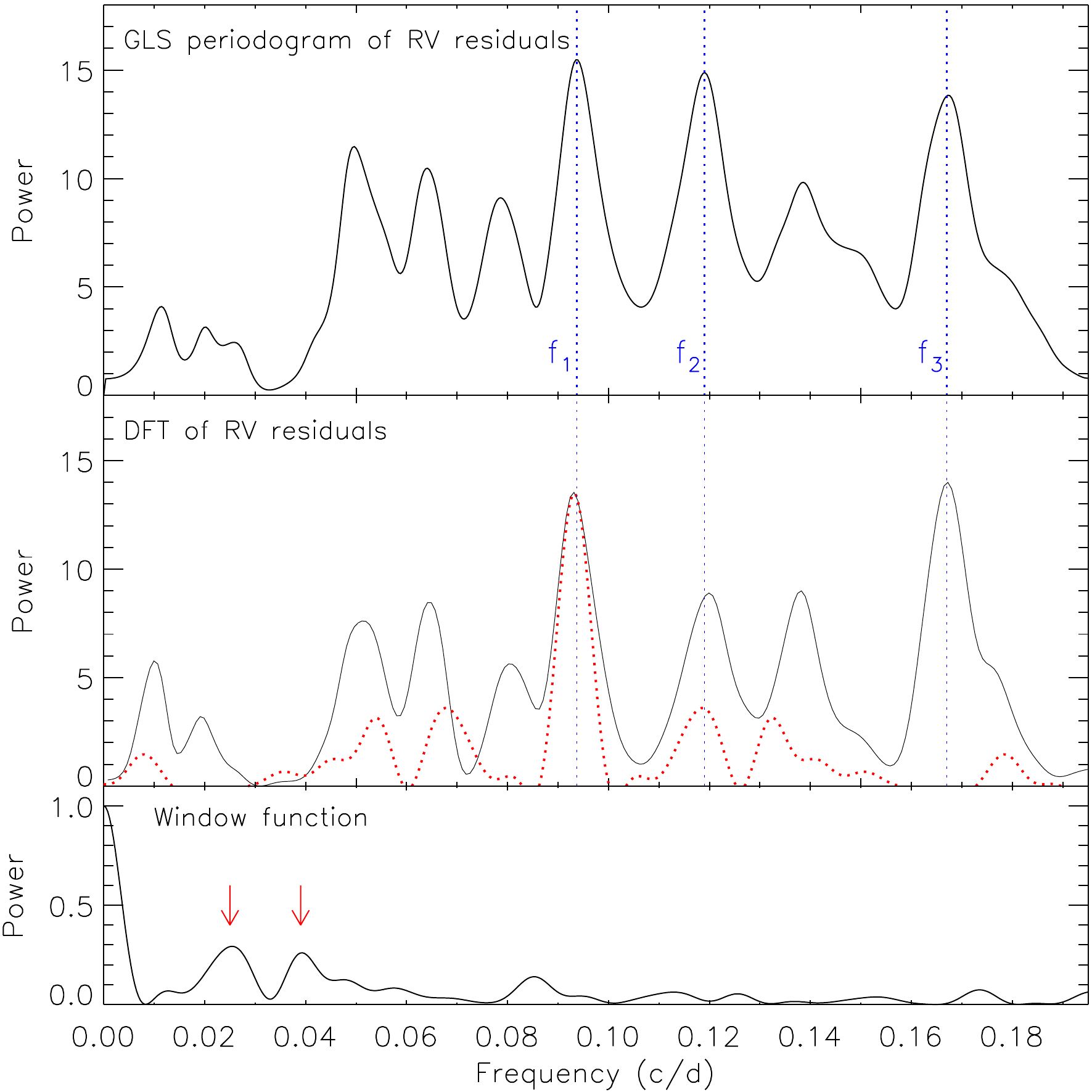}
\caption{\emph{Top panel}: GLS periodograms of the HARPS and HARPS-N RV residuals. The vertical dashed blue lines mark the frequencies $f_1$=0.094~c/d, $f_2$=0.119~c/d, and $f_3$=0.167~c/d whose FAP is less than $10^{-4}$, as derived using a bootstrap randomization procedure. \emph{Middle-panel}: Discrete Fourier transform of the HARPS and HARPS-N RV residuals. The dotted red line marks the window function shifted to the right by $f_1$=0.094~c/d and mirrored to the left of this frequency. \emph{Lower panel}: Window function. The red arrows mark the two peaks presented in the main text.}
\label{GLS_RV_Residuals}
\end{figure}

After fitting the two transiting planets, we inspected the RV residuals to look for additional signals in the Doppler data.~The upper panel of Fig.~\ref{GLS_RV_Residuals} shows the GLS periodogram of the RV residuals (thick black line). There are 3 significant peaks at $f_1$=0.094~c/d ($P_1$=10.7~d), $f_2$=0.119~c/d ($P_2$=8.4~d), and $f_3$=0.167~c/d ($P_3$=6.0~d). We assessed their FAP following the Monte Carlo bootstrap method described in \citet{Kuerster1997}. We computed the GLS periodograms of $10^4$ fake data sets obtained by randomly shuffling the RV measurements, keeping the observation time-stamps fixed. The FAP is defined as the fraction of those periodograms whose highest power exceeds the power spectrum of the original observed data at any frequency.  We found no false positives out of our $10^4$ trials, implying that $f_1$, $f_2$, $f_3$ have a FAP lower than $10^{-4}$.

As a sanity check, we employed the program \texttt{Period04} \citep{Lenz2004} to calculate the discrete Fourier transform (DFT) of the RV residuals. We used the pre-whitening technique \citep[see, e.g.,][]{Hatzes2010} to subsequently identify significant peaks in the power spectrum and remove the corresponding signal from the data. Briefly, we performed a least squares sine-fit to the amplitude and phase at the first dominant frequency found by the DFT and subtracted the fit from the time series. We then reiterated the process to identify and subtract the next dominant Fourier component. The iteration was stopped once we reached the level of the noise. We regarded as significant only those signals whose amplitudes are more than 4 times the Fourier noise level \citep{Breger1993}. The Fourier fit of the RV residuals was obtained with only two dominant frequencies, namely, $f_1$=0.094 c/d and $f_3$=0.167~c/d, with an amplitude of 1.4 and 1.1~\ms, respectively.

The periodogram of the sampling pattern - the so-called ``window function'' - shows two peaks at 0.025~c/d (40~d) and 0.039~c/d (25~d). They are highlighted by two red arrows in the lower panel of Fig.~\ref{GLS_RV_Residuals}. We note that the beat frequency between $f_1$=0.094~c/d and $f_2$=0.119~c/d is equal to 0.025~c/d, which corresponds to one of the two frequencies seen in the window function. This led us to suspect that $f_1$ and $f_2$ are aliases of one another and share the same physical origin. We verified this hypothesis using again the pre-whitening technique. We performed a least-squares sine-fit to the amplitude and phase at either $f_1$ or $f_2$, subtracted the best fit from the RV time series, and recalculated the GLS periodogram of the new residuals. Regardless of which of the two signals is fitted and subtracted first, by removing one of the two we also remove the other, as expected from alias peaks, confirming our hypothesis. We note that the subtraction of the signal at either $f_1$ or $f_2$ does not remove $f_3$=0.167~c/d, which remains significant in the GLS periodogram of the new residuals.

The middle panel of Fig.~\ref{GLS_RV_Residuals} shows the DFT of the RV residuals (thick black line), along with the window function shifted to the right by $f_1$=0.094 c/d and mirrored to the left of this frequency (red dotted line). It is evident that $f_2$, along with most of the side lobes seen to the right and left of $f_1$, is an alias of the latter related to the observing window. We conclude that $f_1$ is very likely the actual periodicity. We also note that $f_3$ is not an alias of $f_1$, as there is no peak detected in the ``shifted'' window function at this frequency, corroborating our pre-whitening analysis.

To further assess which of the two signals is the actual periodicity, we performed a least-squares multi-sine fit to the amplitude and phase at the frequency couples $f_1$, $f_3$, and $f_2$, $f_3$. We then created synthetic RVs residuals using the best fitting parameters, added white noise, sampled the simulated data at the epochs of our real observations, and calculated the GLS periodograms. We found that ``fake'' data sets obtained from the couple $f_1$, $f_3$ reproduce better the observed periodogram than the couple $f_2$, $f_3$. This further supports the fact that the RV residuals contain only two significant signals at $f_1$=0.094~c/d ($P_1$=10.7~d) and $f_3$=0.167~c/d ($P_3$=6.0~d).

What are the sources of the two signals at 6.0 and 10.7 days detected in the RV residuals? Are they due to activity, additional planets, or both? We note that the two periods are close to the first and third harmonic of the rotational period of the star ($P_\mathrm{rot}$\,=\,23.52\,$\pm$\,2.87~days). This might lead us to suspect that magnetic activity coupled with stellar rotation is the source of the two additional signals. Active regions separated by $\sim$90 and $\sim$180 degrees in longitude could account for the two periodicities. To further investigate this hypothesis, we calculated the GLS periodograms of the activity indicators -- namely, the full width at half-maximum (FWHM) and bisector span (BIS) of the cross-correlation profile, and the Ca\,{\sc ii} H\,\&\,K activity index (log\,$R^\prime_\mathrm{HK}$) -- but found no significant peak. We stress, however, that this cannot be used to exclude that activity is the source of the observed RV variation. Given the amplitude of the two signals (1.2 and 1.4 \ms) and low projected rotational velocity of the star (1.8$\pm$0.4\,\kms), the suppression of granular blueshift in magnetized regions of the photosphere of HD\,3167 is expected to be the source of the observed ``jitter''. Based on observations of the Sun as a star, \citet{Haywood2016} recently found that the traditional activity indicators perform poorly in tracing the RV jitter of slowly rotating stars with low level of magnetic activity, such as in the case of HD\,3167.

We further investigated the nature of the additional signals detected in the RV residuals using the stacked Bayesian generalized Lomb-Scargle (BGLS) periodogram proposed by \citet{Mortier2017}. This tool exploits the BGLS algorithm described in \citet{Mortier2015}, which in turn is a Bayesian version of the GLS periodogram of \citet{Zechmeister2009}. As described in \citet{Mortier2017}, the stacked BGLS periodogram relies on the assumption that the power (or probability) of a coherent RV signal -- such as that produced by a \emph{bona fide} orbiting planet -- is expected to increase by adding more data points. On the contrary, the RV signal produced by stellar activity is usually incoherent, since its amplitude, phase, period vary with time, due to the evolution of active regions, differential rotation, and magnetic cycle. Its significance can thus increases or decreases as more RV measurements are added to the data set. The tool calculates the BGLS periodogram for $n$ out of N RVs (where $n\,\le\,N$ ), adds the next point, recalculates the BGLS periodogram, and iterate the process till the last available measurement.

\begin{figure}
\includegraphics[width=\columnwidth]{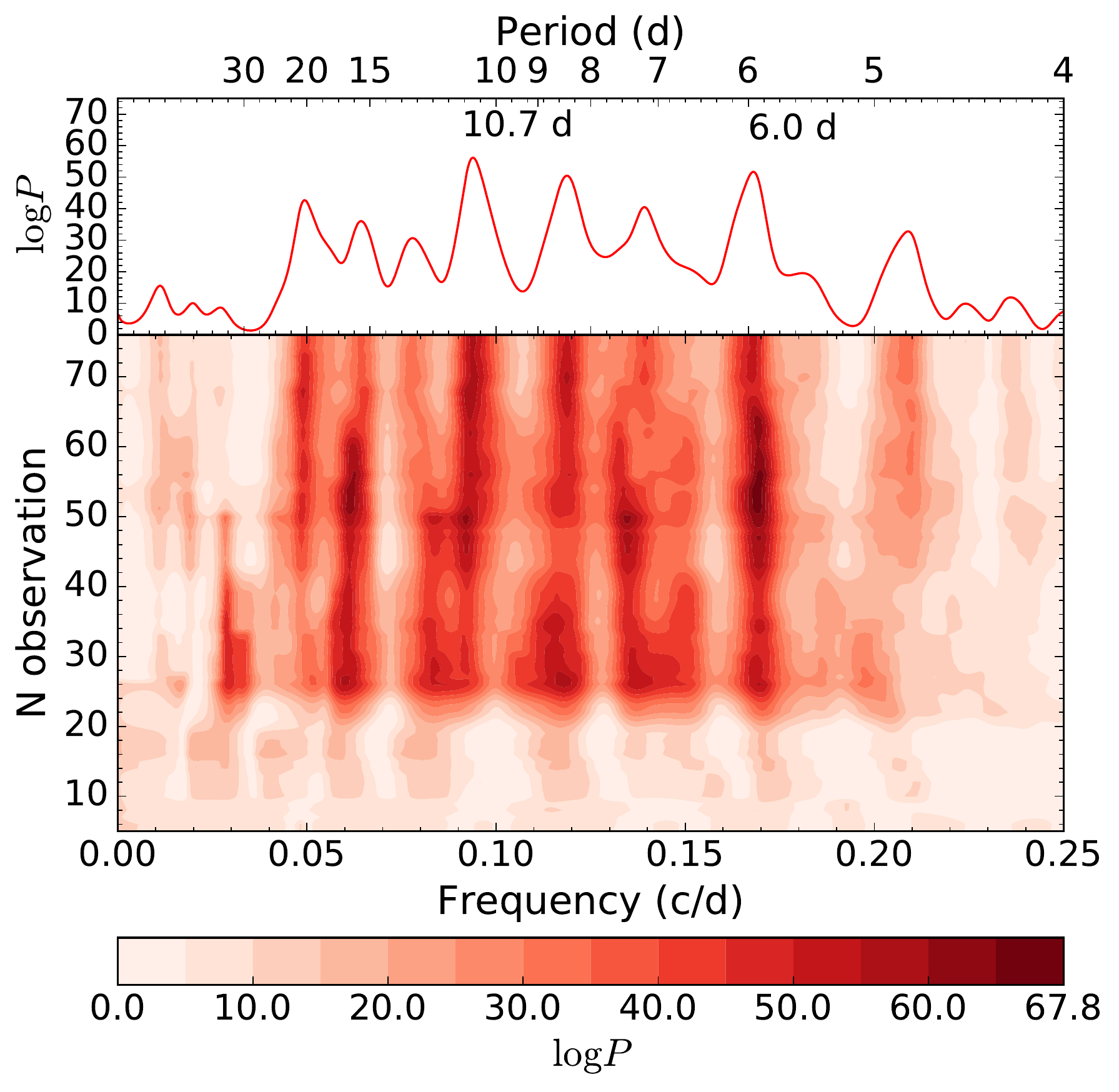}
\caption{Stacked BGLS periodogram of the HARPS and HARPS-N RV residuals.}
\label{SBGLS_RV_Residuals_Periodogram}
\end{figure}

Figure~\ref{SBGLS_RV_Residuals_Periodogram} shows the BGLS periodogram (upper panel) and the stacked BGLS periodogram (lower panel) of the HARPS and HARPS-N RV residuals. As expected, the two dominant peaks at 6.0 and 10.7 days are clearly visible along with their aliases related to the observing window. We note that both signals do not seem to show a steadily increasing power (or probability) as we would expect from signals arising from presence of planets. Is this enough to claim that the two periodic signals are due to activity? Following \citet{Johnson2016}, we created a data set of synthetic RV residuals containing two sinusoidal signals at the same period, phase, and amplitude as the observed data. We added Gaussian noise and sampled the simulated RVs at the time stamps of our observations. 

\begin{figure}
\includegraphics[width=\columnwidth]{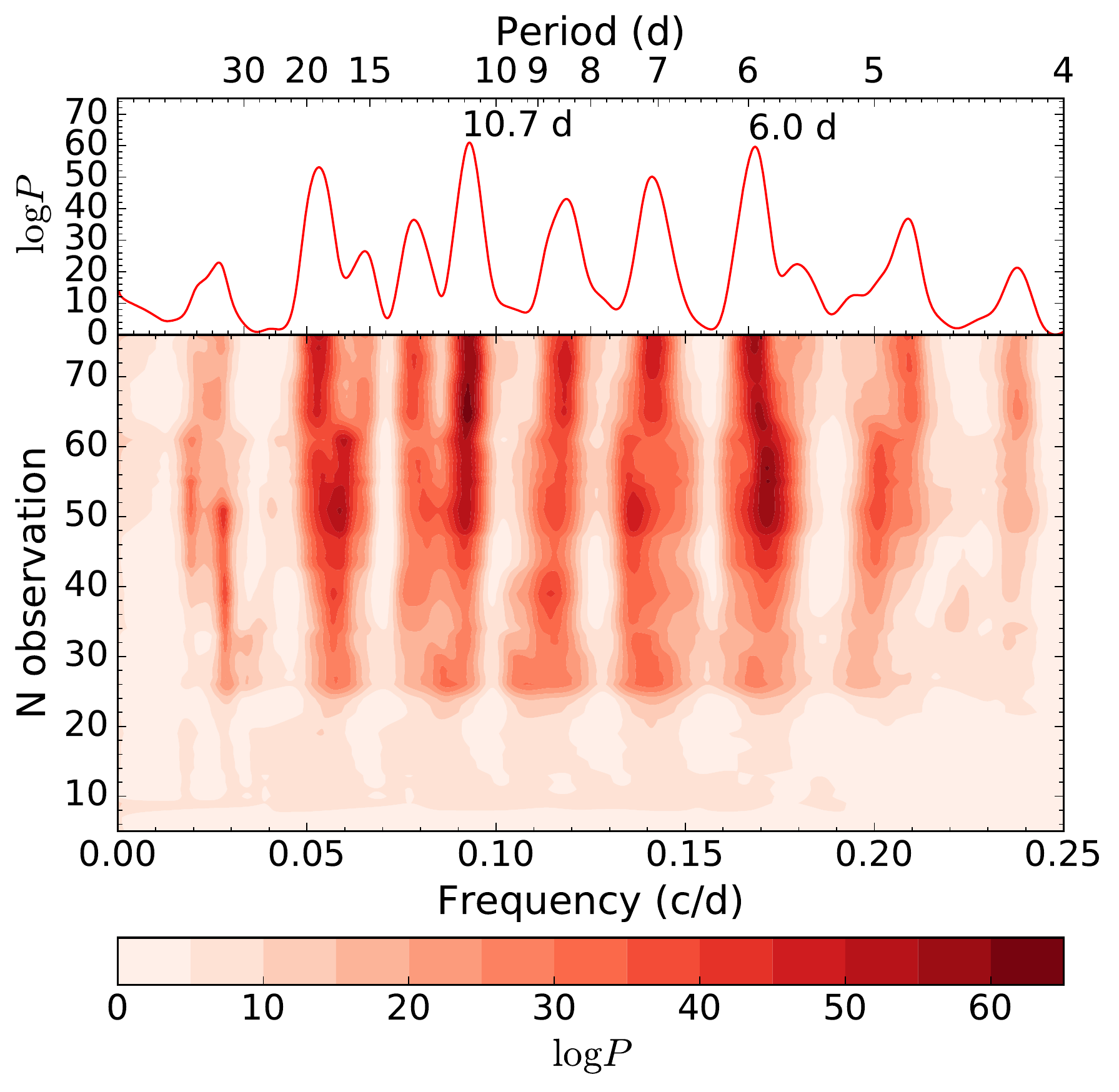}
\caption{Stacked BGLS periodogram of the simulated RV residuals.}
\label{SBGLS_RV_Simulated_Residuals_Periodogram}
\end{figure}

The BGLS periodogram and stacked BGLS periodogram of the synthetic data are shown in Fig.~\ref{SBGLS_RV_Simulated_Residuals_Periodogram}. As is evident from a visual inspection, Fig.~\ref{SBGLS_RV_Residuals_Periodogram} and~\ref{SBGLS_RV_Simulated_Residuals_Periodogram} share roughly the same peaks and a similar pattern. None of the two simulated \emph{coherent} signals shows a steadily increasing power. Given the data, this simulation proves that our sampling of two truly coherent signals at 6.0 and 10.7 days can mimic the trend expected from activity-induced RV variation in the stacked BGLS periodogram.

We conclude that, although we found evidence that there are two additional signals with periods of 6.0 and 10.7~days in the HARPS and HARPS-N measurements, the sampling of our observations, as well as the limited number of RVs and their noise level do not allow us to assess whether the two signals are due to activity, or are rather induced by two additional orbiting planets. We thus include the two signals in our analysis but warn the reader that more observations are needed to unveil their true nature.

\section{Results}
\label{Sect:Results}

We used the code \texttt{pyaneti} to perform the final joint modeling of the \ktwo\ and RV measurements. We fitted the transit and RV curves of HD\,3167\,b and c following the guidelines presented in \S\,\ref{Sect:JointAnalysis}, and incorporated the modeling of the two additional RV signals at 6.0 and 10.7~days using two sinusoidal curves. We set uniform priors for the periods of the two additional signals -- using a 2-day range centered around the values found by the frequency analysis presented in \S\,\ref{Sect:ResidualAnalysis} -- and adopted uninformative priors over a wide range for the corresponding phases and amplitudes. To account for additional instrumental noise not accounted by the nominal RV error bars and/or imperfect treatment of the various sources of RV variations (e.g., stellar activity and/or additional planets), we added jitter terms to the equation of the likelihood for the HARPS and HARPS-N RV data following the method described in \citet{Dumusque2014}. 

\begin{figure*}
\includegraphics[width=\textwidth]{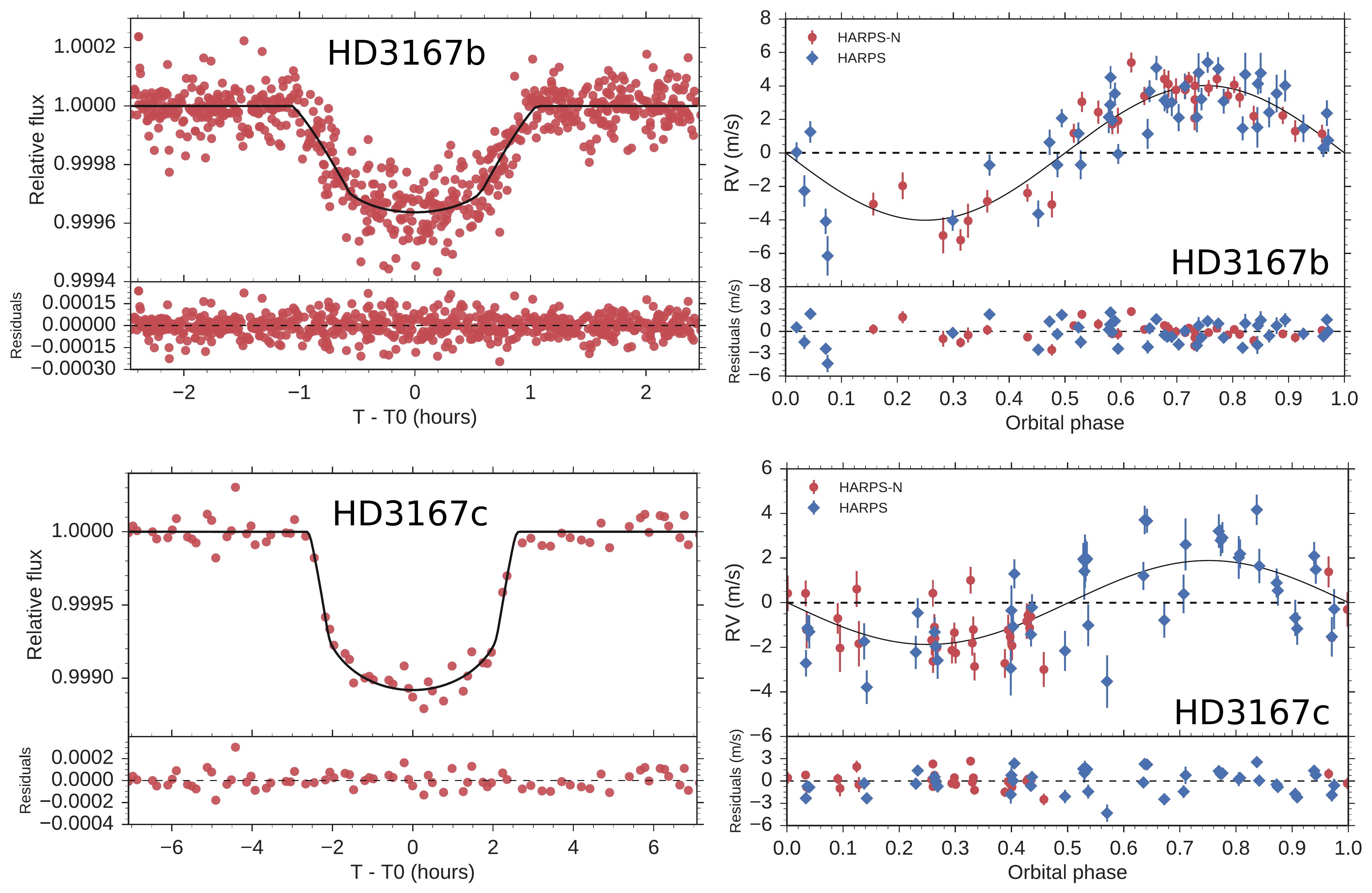}
\caption{ Transit light curves and RV curves of HD\,3167\,b (upper panels) and HD\,3167\,c (lower panels). The best fitting transit and Keplerian models are overplotted with thick black lines. The \ktwo\ data points are shown with red circles (left panels). The HARPS and HARPS-N RV measurements are plotted with red circles and blue diamonds, respectively, along with their nominal uncertainties (right panels).\label{Fig:Joint_Planet_b_c}}
\end{figure*}

\begin{figure}
\includegraphics[width=0.48\textwidth]{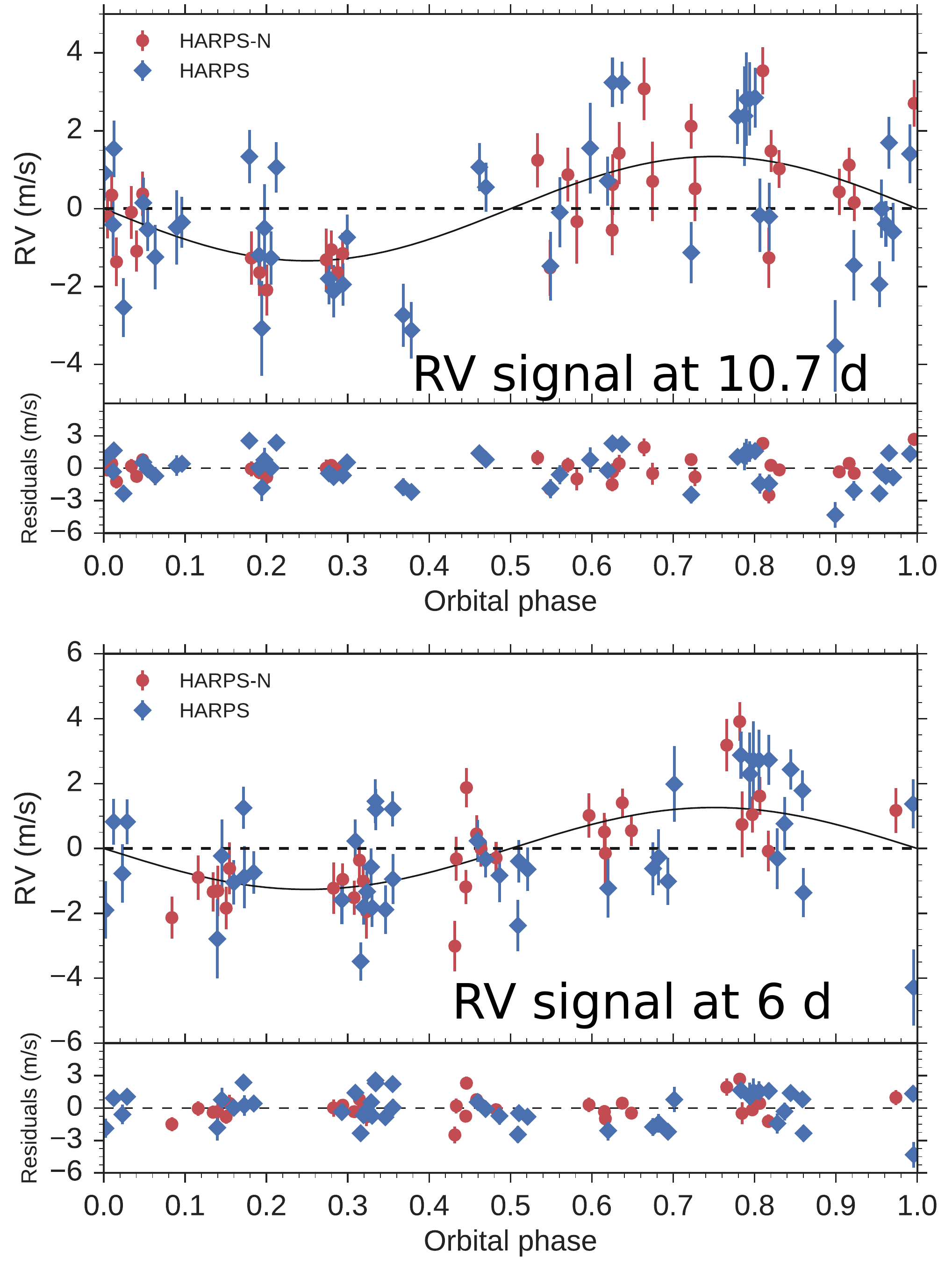}
\caption{Radial velocity curves of the two signals at 10.7 days (upper panel) and 6.0 days (lower panel) and best-fitting models. The HARPS and HARPS-N RV measurements are plotted with red circles and blue diamonds, respectively, along with their nominal uncertainties.}
\label{Fig:Signal_6.0_10.7}
\end{figure}

We report our results in Table~4. The parameter estimates and their error bars were taken to be the median and the 68\,\% credible interval of the final posterior probability distribution of each parameter. Fig.~\ref{Fig:Joint_Planet_b_c} shows the \ktwo\ transit light curves and best fitting transit models, as well as the HARPS and HARPS-N RVs and best fitting Keplerian models of HD\,3167\,b and c. The RV fits to the two additional signals at 6.0 and 10.7 days are shown in Fig.~\ref{Fig:Signal_6.0_10.7}. 

The mass of HD\,3167\,b is in very good agreement with the value we derived using the FCO method corroborating our analysis (cfr. \S\,\ref{HD3167b_FCO}). Similarly, the RV offset between HARPS and HARPS-N ($\Delta$\,RV$_\mathrm{(HS-HN)}$\,=\,8.3\,$\pm$\,0.2\,\ms) agrees with the value presented in \S\,\ref{Sect:GLS_RVs} ($\Delta$\,RV$_\mathrm{(HS-HN)}$\,=\,8.0\,$\pm$\,0.5\,\ms). Finally, our values of the planetary radii agree within less than 1-$\sigma$ with those found by \citet{Vanderburg2016}.

Does the inclusion of the 6.0 and 10.7-day signals bias the mass determinations of HD\,3167\,b~and~HD\,3167\,c\,? A two-planet model fit that included only planet b and c gives RV semi-amplitude variations of $K_{\mathrm b}$\,=\,3.74\,$\pm$\,0.39~\ms\ and $K_{\mathrm c}$\,=\,2.29\,$\pm$\,0.45~\ms, respectively. By adding only the 10.7-day signal we get $K_{\mathrm b}$\,=4.06\,$\pm$\,0.37\,\ms\ and $K_{\mathrm c}$\,=2.04\,$\pm$\,0.43\,\ms. By adding both the 10.7-day and the 6.0-day signal we obtain $K_{\mathrm b}$\,=\,4.02\,$\pm$\,0.31\,\ms\ and $K_{\mathrm c}$\,=\,$1.88^{+0.40}_{-0.42}$\,\ms, proving that the RV semi-amplitude variations -- and thus the determination of the planetary masses of HD\,3167\,b and HD\,3167\,c -- are not significantly affected by the inclusion of the two additional signals.

\section{Discussion and summary}
\label{Sect:Discussion}

\begin{figure*}
\includegraphics[width=\textwidth]{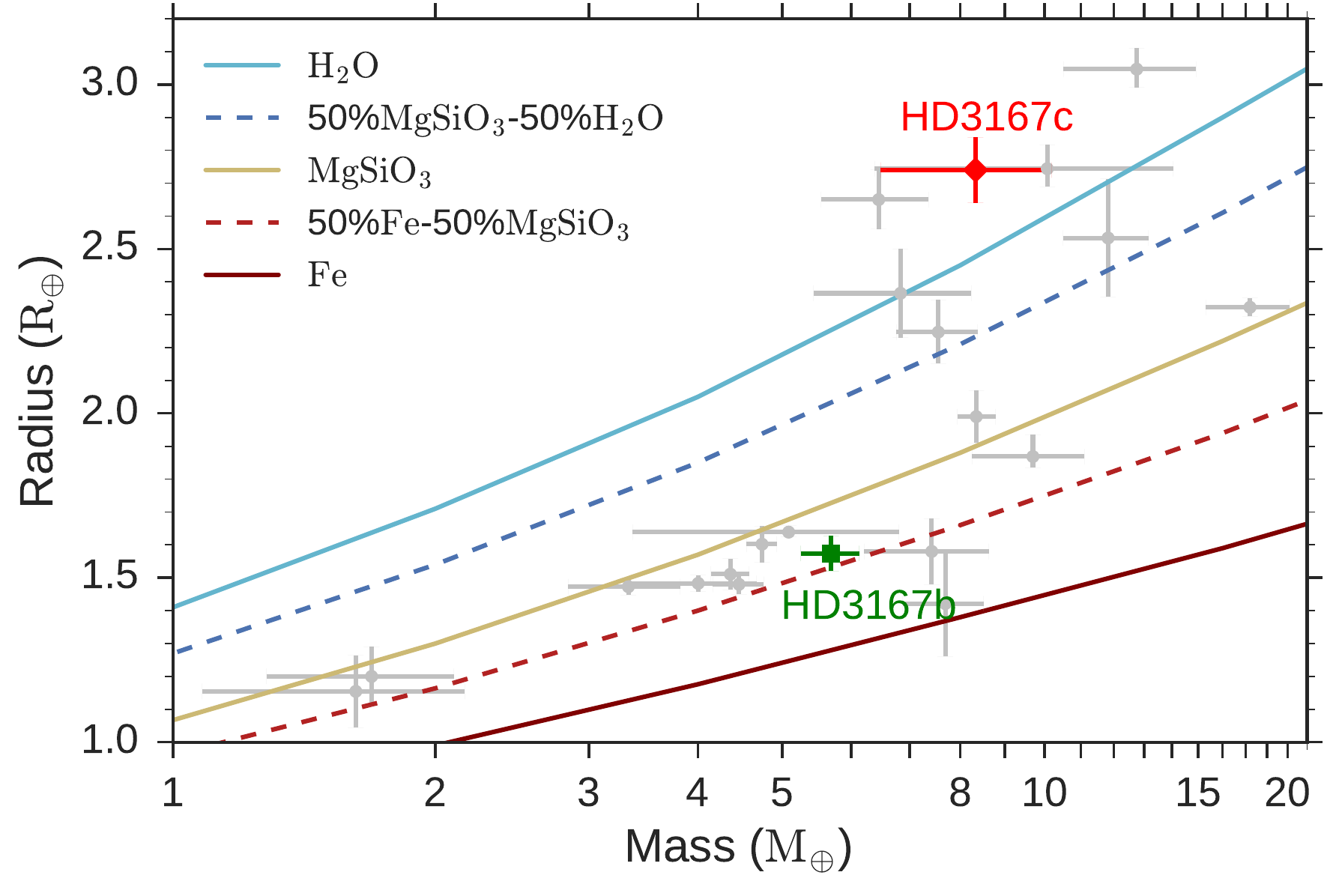}
\caption{Mass-radius diagram for well-characterized (5-$\sigma$ precision level or better) super-Earths and Neptunes. From bottom to top, the solid curves are theoretical models \citep{Zeng2016} for planets with a composition of 100\% iron (brown), 50\% silicate and 50\% iron (dashed red), 100\% silicate (beige), 50\% silicate and 50\% water (dashed blue), water (light blue). HD\,3167\,b \& HD\,3167\,c are highlighted with different symbols and colors.}
\label{Fig:MassRadiusDiagram}
\end{figure*}

The ultra-short period planet HD\,3167\,b has a mass of $M_\mathrm{b}$=\mpb\ and a radius of $R_\mathrm{b}$=\rpb, yielding a mean density of $\rho_\mathrm{b}$=\denpb. Figure~\ref{Fig:MassRadiusDiagram} displays the position of HD\,3167\,b on the mass-radius diagram compared to the sub-sample of small transiting planets ($R\le4$~$R_\oplus$) whose masses and radii have been derived with a precision better than 20\,\%. Theoretical models from \citet{Zeng2016} are overplotted using different lines and colors. The precision of our mass determination (14\,\%) allows us to conclude that HD\,3167\,b is a rocky terrestrial planet with a composition consisting of $\sim$50\,\% silicate and $\sim$50\% iron.

HD\,3167\,b adds to the sample of low-mass, close-in planets with a RV-determined mass and a bulk density suggestive of a mostly rocky composition. Planets belonging to this sample have a restricted Jeans escape parameter $\Lambda  \lesssim 20$ (Table~\ref{tab:planets}). This parameter, defined~as 
\begin{equation}
\Lambda\,=\,\frac{GM_{\rm pl}m_{\rm H}}{k_{\rm B}T_{\rm eq}R_{\mathrm pl}}, 
\end{equation}
has been introduced by \citet{Fossati2017} who found that the hydrogen-dominated atmospheres of exoplanets with $\Lambda \lesssim 20$ lie in the ``boil-off'' regime \citep{Owen2016,Cubillos2017}, where the escape is driven by the atmospheric thermal energy and low planetary gravity. \citet{Fossati2017} also found that the atmosphere of hot ($T_\mathrm{eq}$\,$\gtrsim$\,1000\,K), low-mass ($M_\mathrm{p}$\,$\lesssim$\,5\,$M_\oplus$) planets with $\Lambda \lesssim 20$ shrinks to smaller radii so that their atmosphere evolves out of the ``boil-off'' regime in less than about 500\,Myr.

Because of the very large escape rates after the dispersal of the proto-planetary disc, planets such as HD\,3167\,b have lost quickly (within a few hundreds Myr) their primary hydrogen-dominated atmosphere, as supported, e.g., by the non-detection of a hydrogen exosphere around the ultra-short period planet 55\,Cnc\,e \citep{Ehrenreich2012}. We remark that this fast escape is not driven by the high-energy stellar flux, but by the high temperature of the lower atmosphere and low planetary gravity. This implies that these planets subsequently developed a secondary, possibly CO$_2$-dominated, atmosphere while the host star was still young and hence active. This led to the fast escape -- this time instead driven by the high-energy stellar flux -- also of the secondary atmosphere \citep{Kulikov2006,Tian2009}, leaving behind the strongly irradiated rocky surface. It is therefore foreseeable that the high surface temperature led then to the formation of magma oceans on the day side \citep{Miguel2011,Demory2016}, which out-gases and sputters minerals, forming a tenuous atmosphere not too dissimilar from that of Mercury \citep[e.g.][]{Pfleger2015}. Over time lighter elements escape from the atmosphere, leaving behind a possibly extended exosphere composed mostly by heavy refractory elements that could be detected in transit at ultraviolet and optical wavelengths. This picture would be reinforced if the orbit of HD\,3167\,b had a non-zero eccentricity, as this would lead to tidal heating and thus to a more extended magma ocean. The detection of the exosphere would then enable for the first time the study of the mineralogy of a rocky planet orbiting a star other than the Sun.

With a mass of $M_\mathrm{c}$=\mpc\ and a radius of $R_\mathrm{c}$=\rpc\, the outer planet HD\,3167\,c has a mean density of $\rho_\mathrm{b}$=\denpc, which is consistent with a composition comprising a solid core surrounded by a thick atmosphere. HD\,3167\,c joins the small group of low-density mini-Neptunes with precise mass and radius determinations (Fig.~\ref{Fig:MassRadiusDiagram}).

\begin{table}[t]
\caption{Low-mass ($M$\,$\lesssim$\,8.4\,$M_\oplus$) planets with RV-determined masses, $\Lambda$\,$\lesssim$\,20, and bulk densities suggestive of a mostly rocky composition (mean density $\rho_p > 4$\,g\,cm$^{-3}$). Except for HD\,3167\,b, all values are taken from \citet{Cubillos2017}.}
\label{tab:planets}
\begin{center}
\begin{tabular}{lrc}
\hline
Planet & $\Lambda$ & $\rho_p$     \\
       &           & g\,cm$^{-3}$ \\
\hline
55\,Cnc\,e    & 15.6 & 5.14 \\
CoRoT-7\,b    & 15.6 & 7.97 \\
GJ\,1132\,b   & 18.4 & 5.79 \\
HD\,219134\,b & 20.6 & 5.94 \\
Kepler-10\,b  &  8.9 & 6.31 \\
Kepler-78\,b  &  5.5 & 6.43 \\
Kepler-93\,b  & 18.1 & 6.82 \\
Kepler-97\,b  & 12.3 & 5.93 \\
\hline
HD\,3167\,b   & 15.6 & 8.00 \\
\hline
\end{tabular}
\end{center}
\end{table}

HD\,3167\,c is expected to have a completely different nature with respect to the inner planet b. Despite the lack of mass measurements, \citet{Vanderburg2016} noticed that HD\,3167\,c may be a primary target for transmission spectroscopy. The rather large pressure scale height of about 350\,km and the brightness of the host star (V=8.9 mag) make HD\,3167\,c an ideal target for transmission spectroscopy observations across a wide range of wavelengths, from the far-ultraviolet to the infrared. One can expect the planet to have a rather large hydrogen-rich cloud made of gas escaping from the planetary upper atmosphere under the effect of the high-energy stellar radiation, similarly to GJ\,436\,b \citep{Kulow2014,Ehrenreich2015}. This cloud would be detectable at Ly$\alpha$ during primary transit. Such observations would then provide us with crucial information about the properties of the upper planetary atmosphere and its environment (e.g., stellar wind density and velocity). Observations at longer wavelengths would instead give us the opportunity to study the lower atmosphere and infer its chemical composition and physical properties. HD\,3167\,c appears to be one of the best candidates to investigate the atmosphere of a low-mass planet.

We found evidence for two additional signals with periods of 6.0 and 10.7 days in the HARPS and HARPS-N data. The respective RV semi-amplitude variations are \kd[]\,\ms\ and \ke[]\,\ms. If the signals were caused by two additional orbiting planets, their minimum masses would be \mpd\ and \mpe, respectively. According to the forecasting model of \citet{Chen2017}, the two putative planets would have radii of $\sim$1.9 and 1.5\,$R_{\oplus}$, implying that \ktwo\ would have likely detected their transits if the two planets were transiting HD\,3167. We searched the light curve for additional transit signals using the \texttt{DST} code of \citet{Cabrera2012}, but found none. The null detection of the transits of the two putative additional planets requires that their orbits are inclined by at least $2$-$3$~degrees relative to the orbits of planet b and c. Although a dynamical N-body simulation carried out with \texttt{mercury6} \citep{Chambers1999} shows that such a compact planetary system would be stable for at least $10^7$ years, we stress again that our current data set does not allow us to assess whether the two signals are due planets and/or activity. Additional RV observations are needed to unveil the real nature of the two signals.

\vspace{0.2cm}
We are extremely grateful to the NOT, ESO, TNG staff members for their unique and superb support during the observations. We thank Xavier Bonfils, Fran\c{c}ois Bouchy, Martin K\"urster, Jorge Melendez, and Nuno Santos who kindly agreed to exchange HARPS time with us. D.\,G. would like to acknowledge the inspiring discussions with Conny Konnopke, Nuccio Lanza, Paul Robertson, Rodrigo Diaz, Elisa Delgado Mena, and Aldo Bonomo. D.\,G. gratefully acknowledges the financial support of the \emph{Programma Giovani Ricercatori -- Rita Levi Montalcini -- Rientro dei Cervelli (2012)} awarded by the Italian Ministry of Education, Universities and Research (MIUR). M.\,F. and C.\,M.\,P. acknowledge generous support from the Swedish National Space Board. L.\,F. acknowledges the Austrian Forschungsf\"orderungsgesellschaft FFG project ``TAPAS4CHEOPS'' P853993. Sz.\,C. thanks the Hungarian OTKA Grant K113117. H.\,J.\,D. and D.\,N. acknowledge support by grant ESP2015-65712-C5-4-R of the Spanish Secretary of State for R\& D\&i (MINECO). This research was supported by the Ministerio de Economia y Competitividad under project FIS2012-31079. J.I.G.H and A.S.M. acknowledge financial support from the Spanish Ministry project MINECO AYA2014-56359-P, and J.I.G.H. also from the Spanish MINECO under the 2013 Ram\'on y Cajal program MINECO RYC-2013-14875. The research leading to these results has received funding from the European Union Seventh Framework Programme (FP7/2013-2016) under grant agreement No. 312430 (OPTICON). Based on observations obtained \emph{a}) with the Nordic Optical Telescope (NOT), operated on the island of La Palma jointly by Denmark, Finland, Iceland, Norway, and Sweden, in the Spanish Observatorio del Roque de los Muchachos (ORM) of the Instituto de Astrof\'isica de Canarias (IAC); \emph{b}) with the Italian Telescopio Nazionale Galileo (TNG) also operated at the ORM (IAC) on the island of La Palma by the INAF - Fundaci\'on Galileo Galilei; \emph{c}) the 3.6m ESO telescope at La Silla Observatory under programs ID 097.C-0948 and 098.C-0860. This paper includes data collected by the \kepler\ mission. Funding for the \kepler\ mission is provided by the NASA Science Mission directorate.

\facilities{Kepler (K2), NOT (FIES), ESO-3.6m (HARPS), TNG (HARPS-N)}
\software{\texttt{IDL}, \texttt{SPECTRUM}, \texttt{SME}, \texttt{FAMA}, \texttt{DOOp}, \texttt{PARAM}, \texttt{pyaneti}, \texttt{exotrending}, \texttt{GLS}, \texttt{Period04}}, \texttt{BGLS}, \texttt{DST}, \texttt{mercury6}.

 \floattable
 \begin{deluxetable}{lcc}[t]
 \label{Tab:Parameters}
 \tabletypesize{\scriptsize}
 \tablecolumns{3}
 \tablewidth{0pt}
 \tablecaption{System parameters.}
 \tablehead{
 \colhead{Parameter} &  Prior\tablenotemark{a} & \colhead{Value}}
 \startdata
 \noalign{\smallskip}
  \multicolumn{3}{l}{\emph{\bf{Model Parameters for HD\,3167\,b}}} \\
    Orbital period $P_{\mathrm{orb}}$ (day) &  $\mathcal{U}[0.9596,0.9598]$ & \Pb[] \\
    Transit epoch $T_0$ (BJD$_\mathrm{TDB}-$2\,450\,000) & $\mathcal{U}[7394.3675,7394.3763]$ & \Tzerob[]  \\  
    Scaled semi-major axis $a/R_{\star}$ &  $\mathcal{N}[4.74,0.18]$ & \arb[] \\
    Scaled planet radius $R_\mathrm{p}/R_{\star}$ & $\mathcal{U}[0, 0.5]$ & \rrb[]  \\
    Impact parameter, $b$  & $\mathcal{U}[0,1]$  & \bb[] \\
    Radial velocity semi-amplitude variation $K$ (m s$^{-1}$) & $\mathcal{U}[0,100]$ & \kb[] \\
    $\sqrt{e} \sin \omega$ &  $\mathcal{F}[0]$ & 0 \\
    $\sqrt{e} \cos \omega$ &  $\mathcal{F}[0]$ & 0 \\
   \noalign{\smallskip}
    \multicolumn{3}{l}{\emph{\bf{Derived Parameters for HD\,3167\,b}}} \\
    Planet mass $M_\mathrm{p}$ ($M_{\rm \oplus}$) & $\cdots$ & \mpb[]  \\
    Planet radius $R_\mathrm{p}$ ($R_{\rm \oplus}$) & $\cdots$ & \rpb[] \\
    Mean density  $\rho_\mathrm{b}$ ($\mathrm{g\,cm^{-3}}$) & $\cdots$ & \denpb[]  \\
    Eccentricity $e$ & $\cdots$ & 0 (fixed)  \\
    Semi-major axis of the planetary orbit $a$ (AU) & $\cdots$ & \ab[]  \\
    Orbit inclination  $i_\mathrm{p}$ ($^{\circ}$) & $\cdots$ & \ib[] \\
    Transit duration $\tau_{14}$ (hours) & $\cdots$ & \ttotb[] \\
    Equilibrium temperature$^{(\mathrm{b})}$  $T_\mathrm{eq}$ (K)  & $\cdots$ &  \Tequib[] \\   
    \hline
    \noalign{\smallskip}
    \multicolumn{3}{l}{\emph{\bf{Model Parameters for HD\,3167\,c}}} \\
    Orbital period $P_{\mathrm{orb}}$ (day) &  $\mathcal{U}[ 29.8508 , 29.8532 ]$ & \Pc[] \\
    Transit epoch $T_0$ (BJD$_\mathrm{TDB}-$2\,450\,000) & $\mathcal{U}[ 7394.9763 , 7394.9787 ]$ & \Tzeroc[]  \\  
    Scaled semi-major axis $a/R_{\star}$ &  $\mathcal{N}[ 46.3 , 1.4 ]$ &  \arc[] \\
    Scaled planet radius $R_\mathrm{p}/R_{\star}$ & $\mathcal{U}[0, 0.5]$ & \rrc[]  \\
    Impact parameter, $b$  & $\mathcal{U}[0,1]$  & \bc[]  \\
    Radial velocity semi-amplitude variation $K$ (m s$^{-1}$) & $\mathcal{U}[0,100]$ & \kc[] \\
    $\sqrt{e} \sin \omega$ &  $\mathcal{U}[-1,1]$ & \esinc \\
    $\sqrt{e} \cos \omega$ &  $\mathcal{U}[-1,1]$ & \ecosc \\
    \noalign{\smallskip}
    \multicolumn{3}{l}{\emph{\bf{Derived Parameters for HD\,3167\,c}}} \\
    Planet mass $M_\mathrm{p}$ ($M_{\rm \oplus}$) & $\cdots$ & \mpc []  \\
    Planet radius $R_\mathrm{p}$ ($R_{\rm \oplus}$) & $\cdots$ & \rpc[] \\
    Mean density  $\rho_\mathrm{c}$ ($\mathrm{g\,cm^{-3}}$) & $\cdots$ & \denpc[]\\
    Eccentricity $e$ & $\cdots$ & \ec[]  \\
    Argument of periastron $w_{\star}$ & $\cdots$ & \wc[] \\
    Semi-major axis of the planetary orbit $a$ (AU) & $\cdots$ & \ac[]  \\
    Orbit inclination $i_\mathrm{p}$ ($^{\circ}$) &$\cdots$ & \ic[] \\
    Transit duration $\tau_{14}$ (hours) & $\cdots$ & \ttotc[] \\
    Equilibrium temperature$^{(\mathrm{b})}$  $T_\mathrm{eq}$ (K)  & $\cdots$ &  \Tequic[] \\%
    \hline
    \noalign{\smallskip}
    \multicolumn{3}{l}{\emph{\bf{Signal with period of 10.7\,days}}} \\
    Period $P_{\mathrm{orb}}$ (days) &  $\mathcal{U}[9.4,12.0]$ & \Pd[] \\
    Radial velocity semi-amplitude variation $K$ (\ms) & $\mathcal{U}[0,100]$ & \kd[] \\
    \noalign{\smallskip}
    \multicolumn{3}{l}{\emph{\bf{Signal with period of 6.0\,days}}} \\
    Period $P_{\mathrm{orb}}$ (days) &  $\mathcal{U}[5.4,6.5]$ & \Pe[] \\
    Radial velocity semi-amplitude variation $K$ (\ms) & $\mathcal{U}[0,100]$ & \ke[] \\
    \noalign{\smallskip}
    \multicolumn{3}{l}{\emph{\bf{Other Parameters}}} \\
    Systemic velocity $\gamma_{\mathrm{HARPS}}$  (\kms) & $\mathcal{U}[19.4183,19.6317]$ & \velHARPS[] \\
     Systemic velocity $\gamma_{\mathrm{HARPS-N}}$  (\kms) & $\mathcal{U}[19.4086,19.6197]$ & \velHARPSN[] \\ 
    RV jitter term $\sigma_{\mathrm{HARPS}}$  (\ms) & $\mathcal{U}[0,10]$ & \rvjitterH[] \\   
    RV jitter term $\sigma_{\mathrm{HARPS-N}}$  (\ms) & $\mathcal{U}[0,10]$ & \rvjitterN[] \\
    Parameterized limb-darkening coefficient $q_1$  & $\mathcal{U}[0,1]$ & \qone \\
    Parameterized limb-darkening coefficient $q_2$  & $\mathcal{U}[0,1]$ & \qtwo \\
    Linear limb-darkening coefficient $u_1$ & $\cdots$ & \uone \\
    Quadratic limb-darkening coefficient $u_2$ & $\cdots$ & \utwo\\
   \noalign{\smallskip}
   \enddata
  \tablenotetext{a}{$\mathcal{U}[a,b]$ refers to uniform priors between $a$ and $b$, $\mathcal{N}[a,b]$ to Gaussian priors with mean $a$ and standard deviation $b$, and $\mathcal{F}[a]$ to a fixed $a$ value.}
  \tablenotetext{b}{Assuming zero albedo.}
\end{deluxetable}
\clearpage



\clearpage

\begin{deluxetable*}{lccccccc}
\label{Table:RVs}
\tabletypesize{\normalsize}
\tablecolumns{8}
\tablewidth{0pt}
\tablecaption{FIES, HARPS, HARPS-N radial velocity measurements and activity indicators of HD\,3167.\label{rvs}}
\tablehead{
\colhead{BJD$_\mathrm{TDB}$} & \colhead{RV} & \colhead{$\sigma_{\mathrm{RV}}$} & \colhead{CCF BIS}  & \colhead{CCF FWHM} & \colhead{log\,$R^\prime_\mathrm{HK}$} & \colhead{$\sigma_\mathrm{log\,R^\prime_\mathrm{HK}}$} & \colhead{S/N} per pixel \\
\colhead{$-{2\,450\,000}$} & \colhead{(km s$^{-1}$)} & \colhead{(km s$^{-1}$)}  & \colhead{(km s$^{-1}$)} & \colhead{(km s$^{-1}$)} & \colhead{(dex)} & \colhead{(dex)} & \colhead{@ 5500~\AA} 
}
\startdata
\noalign{\smallskip}
\multicolumn{2}{l}{\bf{FIES}} \\
\noalign{\smallskip}
 7598.642079 &  0.0041 & 0.0024 &      -  &     -  &    -   &    -  &   89.2    \\
 7598.722300 & -0.0016 & 0.0022 &      -  &     -  &    -   &    -  &   94.3    \\
 7599.670737 &  0.0074 & 0.0022 &      -  &     -  &    -   &    -  &   90.2    \\
 \dotfill & \dotfill & \dotfill & \dotfill & \dotfill & \dotfill & \dotfill & \dotfill \\
\noalign{\smallskip}
\hline
\noalign{\smallskip}
\noalign{\smallskip}
\multicolumn{2}{l}{\bf{HARPS}} \\
\noalign{\smallskip}
 7588.842030 & 19.5257 & 0.0009 & -0.0300 & 6.7842 & -5.025 & 0.011 &  82.3 \\
 7589.816345 & 19.5296 & 0.0007 & -0.0279 & 6.7825 & -5.043 & 0.007 &  99.2 \\
 7610.758334 & 19.5197 & 0.0007 & -0.0281 & 6.7784 & -5.075 & 0.009 &  96.9 \\
\dotfill & \dotfill & \dotfill & \dotfill\dotfill & \dotfill & \dotfill & \dotfill & \dotfill \\
\noalign{\smallskip}
\noalign{\smallskip}
\hline
\noalign{\smallskip}
\multicolumn{2}{l}{\bf{HARPS-N}} \\
\noalign{\smallskip}
 7585.641845 & 19.5086 & 0.0006 & -0.0362 & 6.7152 & -5.023 & 0.006 & 113.4 \\
 7587.717619 & 19.5118 & 0.0008 & -0.0377 & 6.7123 & -5.041 & 0.008 &  92.7 \\
 7606.603934 & 19.5119 & 0.0007 & -0.0401 & 6.7100 & -5.045 & 0.007 & 106.7 \\
  \dotfill & \dotfill & \dotfill & \dotfill & \dotfill & \dotfill &\dotfill & \dotfill \\
\enddata
  \tablenotetext{}{The entire RV data set is available in a machine-readable table in the on-line journal.}
\end{deluxetable*}


\begin{thebibliography}{}
\bibitem[Aigrain et al.(2016)]{Aigrain2016} Aigrain, S., Parviainen, H., \& Pope, B.~J.~S.\ 2016, \mnras, 459, 2408 
\bibitem[Baglin \& Fridlund(2006)]{Baglin2006} Baglin, A. \& Fridlund, M. 2006, in "The CoRoT Mission Pre-Launch Status - Stellar Seismology and Planet Finding", ESA-SP 1306, Editors: M. Fridlund, A. Baglin, J. Lochard and L. Conroy. ISBN 92-9092-465-9., p.11
\bibitem[Barrag{\'a}n et al.(2016)]{Barragan2016} Barrag{\'a}n, O., Grziwa, S., Gandolfi, D., et al.\ 2016, \aj, 152, 193 
\bibitem[Barrag{\'a}n et al.(2017)]{Barragan2017} Barrag{\'a}n, O., Gandolfi, D., Smith, A.~M.~S., et al.\ 2017, arXiv:1702.00691
\bibitem[Borucki et al.(2010)]{Borucki2010} Borucki, W.~J., Koch, D., Basri, G., et al.\ 2010, Science, 327, 977 
\bibitem[Breger et al.(1993)]{Breger1993} Breger, M., Stich, J., Garrido, R., et al.\ 1993, \aap, 271, 482 
\bibitem[Bressan et al.(2012)]{Bressan2012} Bressan, A., Marigo, P., Girardi, L., et al.\ 2012, \mnras, 427, 127 
\bibitem[Bruntt et al.(2010)]{Bruntt2010} Bruntt, H.,  Bedding, T.\,R., Quirion, P.-O., et al. 2010, \mnras, 405, 1907
\bibitem[Burke et al.(2015)]{Burke2015} Burke, C.~J., Christiansen, J.~L., Mullally, F., et al.\ 2015, \apj, 809, 8 
\bibitem[Cabrera et al.(2012)]{Cabrera2012} Cabrera, J., Csizmadia, S., Erikson, A., Rauer, H., \& Kirste, S.\ 2012, \aap, 548, A44 
\bibitem[Cantat-Gaudin et al.(2014)]{Cantat2014} Cantat-Gaudin, T., Donati, P., Pancino, E., et al.\ 2014, \aap, 562, A10 
\bibitem[Castelli \& Kurucz(2004)]{Castelli2004} Castelli, F. \& Kurucz, R.\,L. 2004, eprint arXiv: astro-ph/0405087
\bibitem[Chabrier(2001)]{Chabrier2001} Chabrier, G.\ 2001, \apj, 554, 1274 
\bibitem[Chambers(1999)]{Chambers1999} Chambers, J.~E.\ 1999, \mnras, 304, 793 
\bibitem[Chen \& Kipping(2017)]{Chen2017} Chen, J., \& Kipping, D.\ 2017, \apj, 834, 17 
\bibitem[Coelho et al.(2005)]{Coelho2005} Coelho, P., Barbuy, B., Mel\'endez, J., et~al. 2005, \aap, 443, 735
\bibitem[Cosentino et al.(2012)]{Cosentino2012} Cosentino, R., Lovis, C., Pepe, F., et al. 2012, in Society of Photo-Optical Instrumentation Engineers (SPIE) Conference Series, Vol. 8446, Society of Photo-Optical Instrumentation Engineers (SPIE) Conference Series, 1
\bibitem[Csizmadia et al.(2011)]{Csizmadia2011} Csizmadia, S., Moutou, C., Deleuil, M., et al.\ 2011, \aap, 531, A41
\bibitem[Cubillos et al.(2017)]{Cubillos2017} Cubillos, P., Erkaev, N.~V., Juvan, I., et al.\ 2017, \mnras, 466, 1868 
\bibitem[David et al.(2016)]{David2016} David, T.~J., Hillenbrand, L.~A., Petigura, E.~A., et al.\ 2016, \nat, 534, 658 
\bibitem[Demory et al.(2016)]{Demory2016} Demory, B.-O., Gillon, M., de Wit, J., et al.\ 2016, \nat, 532, 207 
\bibitem[Doyle et al.(2014)]{Doyle2014} Doyle, A.\,P., Davies, G.\,R., Smalley, B., et al. 2014, \mnras, 444, 3592
\bibitem[Dumusque et al.(2014)]{Dumusque2014} Dumusque, X., Bonomo, A.~S., Haywood, R.~D., et al.\ 2014, \apj, 789, 154 
\bibitem[Eastman et al.(2013)]{Eastman2013} Eastman, J., Gaudi, B. S.; Agol, E.,  2013, PASP, 125, 83E
\bibitem[Ehrenreich et al.(2012)]{Ehrenreich2012} Ehrenreich, D., Bourrier, V., Bonfils, X., et al.\ 2012, \aap, 547, A18 
\bibitem[Ehrenreich et al.(2015)]{Ehrenreich2015} Ehrenreich, D., Bourrier, V., Wheatley, P.~J., et al.\ 2015, \nat, 522, 		459 
\bibitem[Ford(2006)]{Ford2006} Ford, E.~B.\ 2006, \apj, 642, 505 
\bibitem[Fossati et al.(2013)]{Fossati2013} Fossati, L., Ayres, T.~R., Haswell, C.~A., et al.\ 2013, \apjl, 766, L20 
\bibitem[Fossati et al.(2015)]{Fossati2015} Fossati, L., France, K., Koskinen, T., et al.\ 2015, \apj, 815, 118
\bibitem[Fossati et al.(2017)]{Fossati2017} Fossati, L., Erkaev, N.~V., Lammer, H., et al.\ 2017, \aap, 598, A90
\bibitem[Fridlund et al.(2017)]{Fridlund2017} Fridlund, M., Gaidos, E., Barrag{\'a}n, O., et al.\ 2017, arXiv:1704.08284
\bibitem[Frandsen \& Lindberg(1999)]{Frandsen1999} Frandsen, S. \& Lindberg, B. 1999, in ``Astrophysics with the NOT'', proceedings Eds: Karttunen, H. \& Piirola, V., anot. conf, 71
\bibitem[Gandolfi et al.(2008)]{Gandolfi2008} Gandolfi, D., Alcal{\'a}, J.~M., Leccia, S., et al.\ 2008, \apj, 687, 1303-1322
\bibitem[Gandolfi et al.(2013)]{Gandolfi2013} Gandolfi, D., Parviainen, H., Fridlund, M., et al.\ 2013, \aap, 557, A74 
\bibitem[Gandolfi et al.(2015)]{Gandolfi2015} Gandolfi, D., Parviainen, H., Deeg, H.~J., et al.\ 2015, \aap, 576, A11 
\bibitem[Goodman \& Weare(2010)]{Goodman2010} Goodman, J. \& Weare, J. \ 2010, Comm. App. Math. Comp. Sci., 5, 65
\bibitem[Gray \& Corbally(1994)]{Gray1994} Gray, R.~O., \& Corbally, C.~J.\ 1994, \aj, 107, 742 
\bibitem[Grziwa et al.(2016)]{Grziwa2016} Grziwa, S., Gandolfi, D., Csizmadia, S., et al.\ 2016, \aj, 152, 132
\bibitem[Guenther et al.(2017)]{Guenther2017} Guenther, E.~W., Barragan, O., Dai, F., et al.\ 2017, arXiv:1705.04163 
\bibitem[Gustafsson et al.(2008)]{Gustafsson2008} Gustafsson, B., Edvardsson, B., Eriksson, K., et~al. 2008, \aap, 486, 951
\bibitem[Hatzes et al.(2010)]{Hatzes2010} Hatzes, A.~P., Dvorak, R., Wuchterl, G., et al.\ 2010, \aap, 520, A93 
\bibitem[Hatzes et al.(2011)]{Hatzes2011} Hatzes, A., Fridlund, M., Nachmani, G., et al.\ 2011, \apj, 743, 75
\bibitem[Hatzes(2014)]{Hatzes2014} Hatzes, A.~P.\ 2014, \aap, 568, A84 
\bibitem[Hatzes(2016)]{Hatzes2016} Hatzes, A.~P.\ 2016, \ssr \bibitem[Haywood et al.(2016)]{Haywood2016} Haywood, R.~D., Collier Cameron, A., Unruh, Y.~C., et al.\ 2016, \mnras, 457, 3637 
\bibitem[Heiter et al.(2015)]{Heiter2015} Heiter, U., Lind, K., Asplund, M., et al.\ 2015, \physscr, 90, 054010 
\bibitem[Howard et al.(2009)]{Howard2009} Howard, A.~W., Johnson, J.~A., Marcy, G.~W., et al.\ 2009, \apj, 696, 75 
\bibitem[Howell et al.(2014)]{Howell2014} Howell, S. B.; Sobeck, C.; Haas, M. et al., 2014, PASP..126..398H
\bibitem[Johnson et al.(2016)]{Johnson2016} Johnson, M.~C., Endl, M., Cochran, W.~D., et al.\ 2016, \apj, 821, 74 
\bibitem[K\"urster et al.(1997)]{Kuerster1997} K\"urster, M., Schmitt, J.~H.~M.~M., Cutispoto, G., \& Dennerl, K.\ 1997, \aap, 320, 831 
\bibitem[Kulikov et al.(2006)]{Kulikov2006} Kulikov, Y.~N., Lammer, H., Lichtenegger, H.~I.~M., et al.\ 2006, \planss, 54, 1425 
\bibitem[Kulow et al.(2014)]{Kulow2014} Kulow, J.~R., France, K., Linsky, J., \& Loyd, R.~O.~P.\ 2014, \apj, 			786, 132 
\bibitem[Kurucz(2013)]{Kurucz2013} Kurucz, R.~L.\ 2013, Astrophysics Source Code Library, ascl:1303.024 
\bibitem[L{\'e}ger et al.(2009)]{Leger2009} L{\'e}ger, A., Rouan, D., Schneider, J., et al.\ 2009, \aap, 506, 287
\bibitem[Lenz \& Breger(2004)]{Lenz2004} Lenz, P., \& Breger, M.\ 2004, The A-Star Puzzle, 224, 786 
\bibitem[Lucy \& Sweeney(1971)]{Lucy1971} Lucy, L.\,B. \& Sweeney, M.\,A. 1971, \aj, 76, 544
\bibitem[Lomb(1976)]{Lomb1976} Lomb, N.~R.\ 1976, \apss, 39, 447 
\bibitem[Luger et al.(2016)]{Luger2016} Luger, R., Agol, E., Kruse, E., et al.\ 2016, \aj, 152, 100 
\bibitem[Magrini et al.(2013)]{Magrini2013} Magrini, L., Randich, S., Friel, E., et al.\ 2013, \aap, 558, A38
\bibitem[Mandel \& Agol(2002)]{mandel_2002} Mandel, K. , Agol, E. 2002, ApJ. 580L.171M
\bibitem[Marcy et al.(2014)]{Marcy2014} Marcy, G.~W., Weiss, L.~M., Petigura, E.~A., et al.\ 2014, Proceedings of the National Academy of Science, 111, 12655 
\bibitem[Mayor \& Queloz(1995)]{Mayor1995} Mayor, M., \& Queloz, D.\ 1995, \nat, 378, 355 
\bibitem[Mayor et al.(2003)]{Mayor03} Mayor, M., Pepe, F., Queloz, D., et~al. 2003, Msngr, 114, 20
\bibitem[McQuillan et al.(2014)]{McQuillan2014} McQuillan, A., Mazeh, T., Aigrain, S. 2014, ApJS, 211, 24
\bibitem[Miguel et al.(2011)]{Miguel2011} Miguel, Y., Kaltenegger, L., Fegley, B., \& Schaefer, L.\ 2011, \apjl, 742, L19 
\bibitem[Mortier et al.(2015)]{Mortier2015} Mortier, A., Faria, J.~P., Correia, C.~M., Santerne, A., \& Santos, N.~C.\ 2015, \aap, 573, A101 
\bibitem[Mortier \& Collier Cameron(2017)]{Mortier2017} Mortier, A., \& Collier Cameron, A.\ 2017, arXiv:1702.03885
\bibitem[Nowak et al.(2017)]{Nowak2017} Nowak, G., Palle, E., Gandolfi, D., et al.\ 2017, \aj, 153, 131 
\bibitem[Owen \& Wu(2016)]{Owen2016} Owen, J.~E., \& Wu, Y.\ 2016, \apj, 817, 107
\bibitem[Pepe et al.(2013)]{Pepe2013} Pepe, F., Cameron, A.~C., Latham, D.~W., et al.\ 2013, \nat, 503, 377 
\bibitem[Pfleger et al.(2015)]{Pfleger2015} Pfleger, M., Lichtenegger, H.~I.~M., Wurz, P., et al.\ 2015, \planss, 115, 90 
\bibitem[Queloz et al.(2009)]{Queloz2009} Queloz, D., Bouchy, F., Moutou, C., et al.\ 2009, \aap, 506, 303
\bibitem[Ryabchikova et al.(2011)]{Ryabchikova2011} Ryabchikova, T.~A., Pakhomov, Y.~V., \& Piskunov, N.~E.\ 2011, Kazan Izdatel Kazanskogo Universiteta, 153, 61
\bibitem[Sanchis-Ojeda et al.(2014)]{Sanchis-Ojeda2014} Sanchis-Ojeda, R., Rappaport, S., Winn, J.~N., et al.\ 2014, \apj, 787, 47 
\bibitem[Sanchis-Ojeda et al.(2015)]{Sanchis-Ojeda2015} Sanchis-Ojeda, R., Rappaport, S., Pall{\`e}, E., et al.\ 2015, \apj, 812, 112 
\bibitem[Scargle(1982)]{Scargle1982} Scargle, J.~D.\ 1982, \apj, 263, 835 
\bibitem[Seager \& Mall{\'e}n-Ornelas(2003)]{Seager2003} Seager, S., \& Mall{\'e}n-Ornelas, G.\ 2003, \apj, 585, 1038 
\bibitem[Shkolnik et al.(2014)]{Shkolnik2014} Shkolnik, E.~L., Rolph, K.~A., Peacock, S., \& Barman, T.~S.\ 2014, \apjl, 796, 
\bibitem[Sneden et al.(2012)]{Sneden2012} Sneden, C., Bean, J., Ivans, I., Lucatello, S., \& Sobeck, J.\ 2012, Astrophysics Source Code Library, ascl:1202.009
\bibitem[Stetson \& Pancino(2008)]{Stetson2008} Stetson, P.~B., \& Pancino, E.\ 2008, \pasp, 120, 1332 
\bibitem[Su{\'a}rez Mascare{\~n}o et al.(2015)]{Suarez2015} Su{\'a}rez Mascare{\~n}o, A., Rebolo, R., Gonz{\'a}lez Hern{\'a}ndez, J.~I., \& Esposito, M.\ 2015, \mnras, 452, 2745 
\bibitem[Telting et al.(2014)]{Telting2014} Telting, J.\,H., Avila, G., Buchhave, L., et al. 2014, AN, 335, 41
\bibitem[Tian(2009)]{Tian2009} Tian, F.\ 2009, \apj, 703, 905
\bibitem[Valenti \& Piskunov(1996)]{Valenti1996} Valenti, J.\,A. \& Piskunov, N. 1996, \aaps, 118, 595
\bibitem[Valenti \& Fischer(2005)]{Valenti2005} Valenti, J.\,A. \& Fischer, D.\,A. 2005, \apjs, 159, 141
\bibitem[Vanderburg \& Johnson(2014)]{Vanderburg2014} Vanderburg, A., \& Johnson, J.~A.\ 2014, \pasp, 126, 948 
\bibitem[Vanderburg et al.(2016)]{Vanderburg2016} Vanderburg, A., Bieryla, A., Duev, D.~A., et al.\ 2016, \apjl, 829, L9 
\bibitem[Van Eylen et al.(2016)]{VanEylen2016} Van Eylen, V., Nowak, G., Albrecht, S., et al.\ 2016, \apj, 820, 56 
\bibitem[van Leeuwen(2007)]{vanLeeuwen2007} van Leeuwen, F.\ 2007, \aap, 474, 653
\bibitem[Winn(2010)]{Winn2010} Winn, J.~N.\ 2010, Exoplanets, 55 
\bibitem[Winn \& Fabrycky(2015)]{Winn2015} Winn, J.~N., \& Fabrycky, D.~C.\ 2015, \araa, 53, 409
\bibitem[Zechmeister \& K{\"u}rster(2009)]{Zechmeister2009} Zechmeister, M., \& K{\"u}rster, M.\ 2009, \aap, 496, 577 
\bibitem[Zeng et al.(2016)]{Zeng2016} Zeng, L., Sasselov, D.~D., \& Jacobsen, S.~B.\ 2016, \apj, 819, 127 
\end{thebibliography}
\end{document}